\documentclass[aps,prd,onecolumn,nofootinbib,showkeys]{revtex4-1} 

\usepackage[T1]{fontenc}
\usepackage[utf8]{inputenc}
\usepackage{amsmath,amssymb,amsfonts}
\usepackage{graphicx,epsfig}
\usepackage{color}
\usepackage{hyperref}
\usepackage{array}
\usepackage{subfigure}  
\usepackage{multirow}
\usepackage{float}
\usepackage{booktabs}

\begin{document}

\title{Additional Observational Signatures of Asymmetric Thin-Shell Wormholes within 4D Einstein-Gauss-Bonnet Gravity}

\author{J.SONG}

\author{X.G.Lan}
\email{E-mail: xglan@cwnu.edu.cn}
\affiliation{ Institute of Theoretical Physics, China West Normal University, Nanchong 637009, China}

\date{\today}

\begin{abstract}
In this paper, we study the optical appearance of a 4D Einstein-Gauss-Bonnet asymmetric thin-shell wormhole. Using Visser's cut-and-paste construction, we determine the photon sphere radius and critical impact parameter for different values of the Gauss-Bonnet coupling $\alpha$. We then investigate the effective potential and photon motion in the wormhole spacetime. We find that the effective potential, light-ray trajectories, and azimuthal angle are closely related to the mass ratio of the two spacetimes. Considering an optically thin accretion disk as the only light source, we find that the images of the asymmetric thin-shell wormhole exhibit additional photon rings and lensing bands that are absent for a 4D Einstein-Gauss-Bonnet black hole. Furthermore, the sizes of these extra rings increase with $\alpha$, whereas the photon ring of a 4D EGB black hole shrinks as $\alpha$ increases. This opposite $\alpha$ dependence is the main distinguishing signature of the wormhole geometry, and the resulting exceptionally bright rings may provide a useful observational criterion for distinguishing and characterizing a thin-shell wormhole spacetime. We also verify that the mass ratio and throat radius significantly affect the morphology of these extra photon rings.

\end{abstract}


\maketitle
\tableofcontents  

\section{Introduction}

Unraveling the observational signatures of compact objects is a central research topic in modern strong-field astrophysics. In recent years, astronomical observational techniques have opened two new research avenues that can clarify the intrinsic nature of compact objects and probe the physical behavior of gravitational interactions in strong gravitational fields. 
The first avenue concerns gravitational waves radiated from binary compact object coalescences detected by the LIGO/VIRGO Collaboration, including signals originating from binary black hole mergers and binary neutron star mergers~\cite{Abbott2016a}. The second avenue concerns shadow imaging data of supermassive black holes acquired by the Event Horizon Telescope (EHT)~\cite{Akiyama2019}. These two categories of observational achievements provide solid and compelling evidence for the existence of black holes and the self-consistent validity of General Relativity.

The image of a black hole features a dark central region encircled by a luminous ring, a pattern produced by the interaction of light rays with the black hole's photon sphere (often called the critical curve). The area inside the critical curve is known as the black hole shadow, which naturally lies at the center of the image~\cite{Perlick2022}. The black hole image also encodes information about the jet and surrounding matter, allowing us to probe fundamental properties such as mass, spin, and electric charge. In recent years, the study of black hole shadows in the context of modified theories of gravity has increased rapidly~\cite{Zeng2020a,He2026,Guo2018,Peng2021a,Zeng2020b,Hu2021,Guo2020,Zhong2021,Hou2022,Zeng2022a,Zeng2022b,He2022a,He2022b,Wei2021,Li2021}. For a rotating 4D Einstein-Gauss-Bonnet (EGB) black hole, the deformation of its shadow arises from the spin-dragging effect and is further influenced by the Gauss-Bonnet coupling constant $\alpha$.

Although the EHT results are mainly interpreted within general relativity, they do not entirely rule out the possibility that black holes in modified gravity theories or other ultracompact objects may exist~\cite{Cunha2017,Guo2021,Abdikamalov2019,Narzilloev2020,Herdeiro2021,Rosa2023a,Wang2023}. Indeed, some ultracompact objects might generate shadows that resemble those of black holes. Therefore, there is an urgent need to develop methods for distinguishing black holes from other ultracompact candidates, such as wormholes~\cite{Wielgus2020,Guerrero2021,Tsukamoto2021a,Peng2021b,Guo2023,Chen2023,Nedkova2013,Bugaev2021,Kasuya2021,Bronnikov2021,Tsukamoto2021b,Tsukamoto2022} and boson stars~\cite{Cunha2016,Rosa2022a,Rosa2023b,Rosa2022b}.

In a pioneering work, Wang and colleagues examined the optical appearance of an asymmetric thin-shell wormhole (ATSW) and found that its shadow is smaller than that of a black hole~\cite{Wang2020}. This discovery opened a promising path for the direct observation of wormholes. Later, researchers turned their attention to the double shadows produced by ATSWs, introducing a new radio astronomical technique for distinguishing ultracompact objects from black holes. Other studies have explored the optical appearance and extra photon rings of the Schwarzschild (SS) ATSW~\cite{Peng2021b}, those of an ATSW with a Hayward profile~\cite{Guo2023}, and the optical appearance of a star falling into an ATSW~\cite{Chen2023}. Olmo et al. investigated the observational appearance of black holes and traversable wormholes, along with the Lyapunov exponents of unstable circular photon orbits~\cite{Olmo2023}. Recently, Macedo et al. found that photon scattering off the wormhole throat yields a peculiar shadow structure, which deviates significantly from the shadow silhouette of a black hole in a certain parameter space~\cite{Macedo2026}.

It is widely recognized that, under the assumptions of the strong energy condition and global hyperbolicity, a black hole in general relativity must contain a singularity \cite{Hawking1970}. This singularity is generally regarded as a symptom of the incompleteness of general relativity, and it may be resolved by quantum-gravity corrections. In this context, the Gauss-Bonnet coupling constant $\alpha$ opens up the possibility of constructing four-dimensional regular black holes, providing a theoretical foundation for the asymmetric thin-shell wormhole model studied here.

Previous studies have demonstrated that the Gauss-Bonnet coupling constant $\alpha$ affects the shadow properties of a black hole: for a 4D EGB black hole surrounded by a thin accretion disk, a larger $\alpha$ reduces both the size of the shadow and the size of its surrounding bright rings~\cite{KumarGhosh2020}. In the present paper, we focus on the optical appearance of an asymmetric thin-shell wormhole endowed with a 4D EGB profile. On the one hand, because the observational appearance of the Schwarzschild ATSW has already been studied, we are motivated to investigate whether variations in $\alpha$ induce analogous changes in the observational features of the 4D EGB ATSW, similar to those found for the 4D EGB black hole. On the other hand, we examine whether the shadow characteristics could serve as a possible basis for distinguishing the ATSW from a black hole, thereby offering a potentially useful observational approach to studying strong-gravity systems.

It is worth clarifying what our work adds beyond existing ATSW studies. The Schwarzschild ATSW is the limiting case of our model as $\alpha \to 0$~\cite{Peng2021b}. Once a nonzero $\alpha$ is introduced, the positions and separations of the extra rings shift systematically, and the direction of the shift depends directly on the sign and magnitude of $\alpha$. For the Hayward ATSW, the relevant parameters, such as the magnetic charge $g$, mainly modify the spacetime structure near the horizon, and their influence on the strong-lensing region outside the photon sphere is relatively local~\cite{Guo2023}. In contrast, the 4D EGB coupling $\alpha$ enters directly into the square-root term of the metric function $f_i(r)$ and simultaneously changes the photon-sphere radius, the critical impact parameter, and the ratio of the impact parameters on the two sides, $Z$. It therefore produces a more global control over the distribution of the extra photon rings and lensing bands. In addition, Ref.~\cite{Tan2026} studied the observational features of an ATSW in the Kalb-Ramond background and found that, as $Q$ and $l$ increase, the coverage area of a specific extra halo expands accordingly. The novelty of our work lies in systematically studying, for the first time, how $\alpha$ regulates the extra rings of an ATSW. Here, $\alpha$ is a purely gravitational-theory parameter, and we find that its trend is opposite to that in the black hole case in the same theory.

This work is organized as follows. In Section II, we construct a 4D EGB asymmetric thin-shell wormhole and analyze its geodesics, followed by calculations of photon trajectories and light deflection in this wormhole background. In Section III, we investigate the transfer functions and the observational appearances of both the black hole and the asymmetric thin-shell wormhole under two different emission models. In Section IV, we present our conclusions and further thoughts on the prospects for the observational detectability of these wormholes. Throughout this paper, we adopt units with $G=c=1$.

\section{Effective Potential and Null Geodesic of The Asymmetric Thin-Shell Wormhole with a 4D EGB Profile}\label{sec2}

Before proceeding, we specify the theoretical framework adopted in this work. In spacetime dimensions $D<5$, the Gauss--Bonnet Lagrangian 
\begin{equation}
\mathcal{L}_{\mathrm{GB}} = R^{\mu\nu\rho\sigma}R_{\mu\nu\rho\sigma}
- 4R^{\mu\nu}R_{\mu\nu} + R^{2}
\end{equation}
is a total derivative and therefore does not contribute to gravitational dynamics. However, Tomozawa pointed out that, at the quantum level, a consistent regularization procedure can yield a nontrivial contribution from the Gauss--Bonnet term in a four-dimensional static, spherically symmetric black-hole spacetime \cite{Tomozawa2012}, with the metric
\begin{equation}
ds^{2} = -f(r)dt^{2} + \frac{1}{f(r)}dr^{2}
+ r^{2}\left(d\theta^{2} + \sin^{2}\theta\,d\phi^{2}\right),
\end{equation}
where
\begin{equation}
f(r) = 1 + \frac{r^{2}}{32\pi\alpha}
\left[1 \pm \sqrt{1 + \frac{128\pi\alpha M}{r^{3}}}\right].
\end{equation}
We choose the negative branch. Subsequently, Glavan and Lin proposed
rescaling the Gauss-Bonnet coupling parameter as
$\alpha \to \alpha/(D-4)$ and taking the limit $D \to 4$ at the level
of the field equations, thereby allowing the Gauss-Bonnet term to
produce nontrivial dynamical effects \cite{Glavan2020}. This theory
contains only the massless graviton and is free from ghost
instabilities, as in general relativity. The 4D EGB metric function
employed in this work is precisely the static, spherically symmetric
solution obtained within this framework.

It should be noted that the theoretical status of 4D EGB gravity has
been discussed in the literature, and the well-definedness of the
rescaled field equations in four dimensions remains under debate. We
emphasize, however, that all calculations in this work---the
photon-sphere radius, the critical impact parameter, the effective
potential, the null geodesics, the deflection angle, and the transfer
functions---depend only on the specific form of the spherically
symmetric metric function $f(r)$, and not on the well-definedness of
the four-dimensional field equations. Moreover, the same spherically
symmetric metric also arises in a variety of frameworks, including the
semiclassical conformal anomaly, quantum corrections, regularized
Lovelock gravity, and Horndeski scalar-tensor theory
\cite{KumarGhosh2020}. The background solution adopted here is
therefore robust across different regularization schemes. Accordingly,
the scope of our results is limited to the observational phenomenology
based on this spherically symmetric metric.

In this work, $\alpha$ denotes the Gauss-Bonnet coupling constant. It
controls the strength of the quadratic curvature correction, and its
sign and magnitude affect the photon-sphere radius, the critical impact
parameter, and the ratio of the impact parameters on the two sides,
$Z$, thereby regulating the extra photon rings of the wormhole. We
adopt units with $G=c=1$ and set $M_1=1$, so that $\alpha$ is measured
in units of $M_1^2$. The range $\alpha \in [-0.3, 0.3]$ used here
corresponds to a weak-coupling regime; for these values, the metric
function $f_i(r)$ remains real and the shell lies outside both event
horizons. Observational constraints on $\alpha$ from black-hole shadows
are relatively weak, so the values adopted here are physically
reasonable for an initial exploration.

In addition, we construct the asymmetric thin-shell wormhole using
Visser's cut-and-paste method, in which two 4D EGB spherically
symmetric spacetimes are joined along a thin shell. This construction
requires only that the metric be continuous across the shell,
$g_{\mu\nu}^{+}(R) = g_{\mu\nu}^{-}(R)$, and we adopt
$R > \max(r_{h1}, r_{h2})$ to ensure that the shell lies outside both
event horizons. Because the construction relies only on the metric
function itself and does not involve the well-definedness of the
four-dimensional field equations, the theoretical debate discussed
above does not affect the validity of the thin-shell construction in
this work. The choice $\alpha=\pm 0.3$ is made to explore the
qualitative effect of $\alpha$ on the extra rings; it is not intended
to represent a specific observational bound. The sign of $\alpha$ is
kept free so that both cases can be compared, and its magnitude is
chosen to be small enough for the weak-coupling treatment to remain
valid.

We use Visser's cut-and-paste method to construct an ATSW with a 4D EGB profile~\cite{Visser1989}. The ATSW consists of two spacetimes $\mathcal{M}_1$ and $\mathcal{M}_2$ with different mass parameters $M_1$ and $M_2$, connected by a thin-shell throat. The ATSW can be represented by the manifold $\mathcal{M} \equiv \mathcal{M}_1 \cup \mathcal{M}_2$, with the metric
\begin{equation}\label{eq1}
\begin{aligned}
\mathrm{d}s_i^2 = -f_i(r_i)\,\mathrm{d}t_i^2 + \frac{1}{f_i(r_i)}\,\mathrm{d}r_i^2 + r_i^2\left(\mathrm{d}\theta_i^2 + \sin^2\theta_i\,\mathrm{d}\phi_i^2\right), 
\end{aligned}
\end{equation}
where the metric functions are specified by the 4D EGB black hole solution~\cite{Glavan2020}.
\begin{equation}\label{eq2}
\begin{gathered}
f_i(r_i) = 1 + \frac{r_i^2}{2\alpha} \left(1 - \sqrt{1 + \frac{8\alpha M_i}{r_i^3}}\right), \qquad r_i \geq R, 
\end{gathered}
\end{equation}
where $M_i$ denotes the mass parameter and $\alpha$ is the dimensionless Gauss--Bonnet coupling constant. $R$ denotes the position of the thin shell and satisfies $R > \max(r_{h1}, r_{h2})$, where $r_{h1}$ and $r_{h2}$ are the event horizon radii of the corresponding black holes. Taylor expanding Eq.~(5) and taking the limit $\alpha \to 0$, the metric in Eq.~(4) reduces to the Schwarzschild metric.

Before applying the constancy of photon four-momentum across the
shell, we establish this property more rigorously within the thin-shell
construction. We follow the general framework of Garcia, Lobo, and
Visser \cite{GarciaLoboVisser2012}, adapted to our four-dimensional,
spherically symmetric EGB background.

The two spacetimes are joined along the shell, whose induced metric is
$h_{ab} = g_{\mu\nu} e_{(a)}^{\mu} e_{(b)}^{\nu}$. For the static,
spherically symmetric background considered here, this metric can be written as
\begin{equation}
ds_{\Sigma}^{2} = -d\tau^{2} + R^{2}(\tau)\left(d\theta^{2}
+ \sin^{2}\theta\, d\phi^{2}\right),
\end{equation}
where $\tau$ is the proper time of an observer on the shell and
$R(\tau)$ is the shell radius. The $\theta$ and $\phi$ directions are
automatically continuous, whereas the time coordinates on the two sides
are generally not identical. They must therefore be chosen such that the
proper time on the shell is the same on both sides. This gives the rescaling
\begin{equation}
dt_{2} = \sqrt{\frac{f_{1}(R)}{f_{2}(R)}}\, dt_{1}.
\end{equation}

The tangential components of the photon four-momentum are continuous across the shell. Therefore, the angular momentum is the same on both sides, $L_{1} = L_{2}$, because the two sides share the same $\phi$ coordinate and the shell is spherically symmetric. For the energy component, the rescaling of the time coordinate implies that the conserved energies $E_{i} = -p_{t_{i}}$ on the two sides satisfy $E_{2}/E_{1} = \sqrt{f_{1}(R)/f_{2}(R)}$. For the parameters used in this work ($M_{1}=1$, $M_{2}=1.2$, $R=2.6$, and $\alpha \in [-0.3, 0.3]$), we have verified numerically that $\sqrt{f_{1}(R)/f_{2}(R)}$ is close to $1$. Thus, the conserved energies on the two sides can be treated as approximately equal. Strictly speaking, the ratio is given by $Z = \sqrt{f_{1}(R)/f_{2}(R)}$; however, for our parameters, $Z$ is already very close to $1$, and the approximation of equal energies is standard in the thin-shell wormhole literature.

The surface stress tensor on the shell follows from the Israel junction conditions (the Lanczos equations)~\cite{GarciaLoboVisser2012}. For a spherically symmetric shell, it is described by a surface energy density and a surface pressure. As pointed out in that work, the surface energy density is always negative, implying that the null and weak energy conditions are violated on the shell. This is consistent with the well-known result for traversable wormholes: exotic matter that violates the energy conditions is required to keep the throat open. Since this work focuses on the optical appearance rather than the physical origin of the shell matter, we do not further examine the nature of this exotic matter.

The stability condition for the throat of an asymmetric Schwarzschild thin-shell wormhole is:
\begin{equation}
R\, m_{s}''(R) \geq F(M_{1}, M_{2}, R),
\end{equation}
where
\begin{equation}
F(M_{1}, M_{2}, R) = \frac{(M_{1}/R)^{2}}{(1 - 2M_{1}/R)^{3/2}}
+ \frac{(M_{2}/R)^{2}}{(1 - 2M_{2}/R)^{3/2}}.
\end{equation}
Here, $R$ denotes the shell radius in our work, $M_{1}$ and $M_{2}$ are the mass parameters on the two sides, and $m_{s}(R)$ is the surface mass of the shell. This inequality indicates that stability is controlled jointly by the two masses and the shell radius, and that the stable region lies above the surface $F(M_{1}, M_{2}, R)$.

It is worth noting that Eq.~(9) also uses a stability condition, namely $V''(r_{ph_{i}}) < 0$, to determine the photon sphere (i.e.\ the maximum of the effective potential). The stability condition in Eq.~(9) concerns the stability of circular photon orbits, whereas the Garcia-Lobo-Visser condition concerns the dynamical stability of the throat under radial perturbations. These are two distinct conditions, although both fall under the general heading of ``stability conditions''.

Considering that the gravitational interaction is the only force acting on a photon as it passes through the thin shell, the four-momentum $p^a$ of the photon remains constant. In spacetime $\mathcal{M}$, we have $g_{\mu\nu}^{\mathcal{M}_1}(R) = g_{\mu\nu}^{\mathcal{M}_2}(R)$ because of the continuity of the metric~\cite{Visser1989}. This system has two conserved quantities, namely $p_{t_i} = -E_i$ and $p_{\phi_i} = L_i$, when the photon moves along a geodesic and satisfies the equations of motion.
\begin{equation}
\frac{p_{t_i}^2}{f_i(r_i)} - \frac{p_{\phi_i}^2}{r_i^2} - \frac{(p_i^{r_i})^2}{f_i(r_i)} = 0 .
\end{equation}
In Eq.~(10), $p_i^{r_i} = dr_i/d\lambda$ denotes the radial component of the photon's four-momentum, where $\lambda$ is an affine parameter. Rearranging Eq.~(10) gives the following expression:
\begin{equation}
p_i^{r_i} = \pm E_i \sqrt{1 - \frac{b_i^2}{r_i^2} f_i(r_i)} .
\end{equation}
The plus and minus signs denote outgoing and incoming photons, respectively. The impact parameter is defined as $b_i = |L_i| / E_i$. Substituting Eq.~(11) into the radial equation of motion yields the effective potential $V_i(r_i)$ for the 4D EGB wormhole:
\begin{equation}
V_i(r_i) = \frac{f_i(r_i)}{r_i^2} = \frac{1}{r_i^2} + \frac{1}{2\alpha}\left(1 - \sqrt{1 + \frac{8\alpha M_i}{r_i^3}}\right) .
\end{equation}
For photons on unstable circular orbits (i.e., the photon sphere), the corresponding conditions are as follows:
\begin{equation}
V_i(r_{ph_i}) = \frac{1}{b_{c_i}^2}, \quad V_i'(r_{ph_i}) = 0, \quad V_i''(r_{ph_i}) < 0.
\end{equation}

Here, $b_{c_i}$ and $r_{ph_i}$ denote the critical impact parameter and the photon sphere radius, respectively. When the equality $b = b_c$ is satisfied, the light ray has an impact parameter arbitrarily close to the critical value associated with the photon sphere, causing the ray to circle the black hole infinitely many times. The stability of this orbit is determined by the sign of the second derivative. The unstable circular orbit that defines the photon sphere corresponds to a local maximum of the effective potential. Combining Eqs.~(12) and~(13) yields the expressions for $r_{ph_i}$ and $b_{c_i}$:

\begin{equation}
r_{ph_i} = 2\sqrt{3}M_i \cos\left(\frac{1}{3}\arccos\left(-\frac{4\alpha}{3\sqrt{3}M_i^2}\right)\right),
\qquad
b_{c_i} = \frac{r_{ph_i}}{\sqrt{f_i(r_{ph_i})}},
\end{equation}
with
\begin{equation}
f_i(r_i)=1+\frac{r_i^2}{2\alpha}\left(1-\sqrt{1+\frac{8\alpha M_i}{r_i^3}}\right).
\end{equation}

Consequently, $S_{p_i}$ denotes the area of the region enclosed by the photon sphere in the two-dimensional plane and is given by $S_{p_i} = \pi r_{ph_i}^2$.

\begin{figure}[htbp]
\centering
\begin{minipage}[b]{0.3\textwidth}
    \centering
    \includegraphics[width=\linewidth]{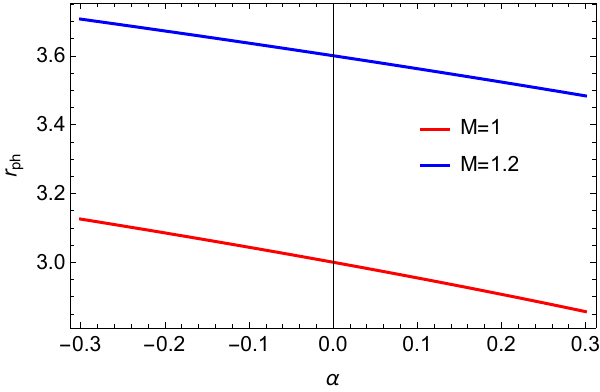}
    \small (a) Photon sphere radius $r_{ph}$.
\end{minipage}
\hfill
\begin{minipage}[b]{0.3\textwidth}
    \centering
    \includegraphics[width=\linewidth]{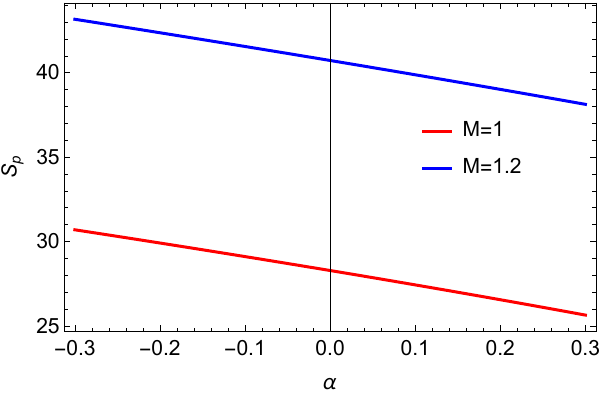}
    \small (b) Enclosed area $S_{p}=\pi r_{ph}^2$.
\end{minipage}
\hfill
\begin{minipage}[b]{0.3\textwidth}
    \centering
    \includegraphics[width=\linewidth]{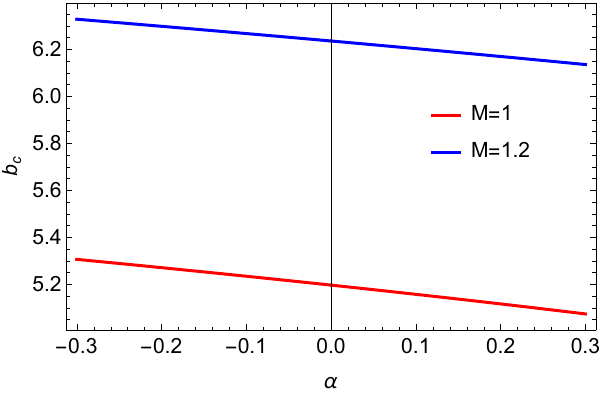}
    \small (c) Critical impact parameter $b_c$.
\end{minipage}
\caption{(Color online) Photon sphere radius $r_{ph}$, enclosed area $S_{\pi}=\pi r_{ph}^2$, and critical impact parameter $b_c$ as functions of the Gauss--Bonnet coupling constant $\alpha$ for $M=1$ (red curves) and $M=1.2$ (blue curves). The range of $\alpha$ is $\alpha\in[-0.3,0.3]$.}
\label{fig:parameters}
\end{figure}

When the photon sphere lies outside the event horizon, distinguishing the wormhole from a black hole becomes a nontrivial task. We assume that the observer resides in spacetime $\mathcal{M}_1$, with mass parameter $M_1 = 1$. For the opposite side, we denote the mass parameter as $M_2 = k$. The parameters $k$ and the throat radius $R$ must satisfy the following inequality~\cite{Wang2020,Peng2021b}:
\begin{equation}
1 < k < \frac{R}{2} \leq \frac{r_{ph_1}}{2}.
\end{equation}

In practice, any choice of $k$ satisfying $1 < k < R/2$ yields qualitatively similar results. Therefore, we fix $M_2 = k = 1.2$. For the two distinct spacetimes $\mathcal{M}_1$ and $\mathcal{M}_2$, the impact parameters on either side are related by the following expression~\cite{Wang2020}:
\begin{equation}
\frac{b_1}{b_2} = \sqrt{\frac{f_2(R)}{f_1(R)}} = \sqrt{\frac{1 + \frac{R^2}{2\alpha}\left(1 - \sqrt{1 + \frac{8\alpha M_2}{R^3}}\right)}{1 + \frac{R^2}{2\alpha}\left(1 - \sqrt{1 + \frac{8\alpha M_1}{R^3}}\right)}} \equiv Z. 
\end{equation}
With $M_1 = 1$ and $M_2 = 1.2$ fixed, we solve Eq.~(13) numerically to obtain the photon sphere radii $r_{ph_i}$ and the corresponding critical impact parameters $b_{c_i}$ for various values of the Gauss--Bonnet coupling constant $\alpha$. The results are summarized in Tab.~I. Increasing $\alpha$ reduces both $r_{ph_i}$ and $b_{c_i}$, indicating that a larger $\alpha$ shifts the photon sphere inward. In the limit $\alpha \to 0$, the spacetime reduces to the Schwarzschild ATSW.
\begin{table}[htbp]
\centering
\caption{Critical impact parameters and photon sphere radii for the 4D EGB asymmetric thin-shell wormhole. Different values of the Gauss--Bonnet coupling constant $\alpha$ are considered. For spacetime $\mathcal{M}_1$, $M_1 = 1$; for spacetime $\mathcal{M}_2$, $M_2 = 1.2$.}
\begin{tabular}{ccccc}
\toprule
$\alpha$ & $b_{c1}$ & $r_{ph1}$ & $b_{c2}$ & $r_{ph2}$ \\
\midrule
-0.3 & 5.30178 & 3.11868 & 6.32565 & 3.70216 \\
-0.2 & 5.26854 & 3.08198 & 6.29679 & 3.66992 \\
SS & 5.19615 & 3.00000 & 6.23538 & 3.60000 \\
0.2  & 5.11331 & 2.90178 & 6.16799 & 3.52081 \\
0.3  & 5.06641 & 2.84340 & 6.13148 & 3.47664 \\
\bottomrule
\end{tabular}
\label{tab:params}
\end{table}
Fig.~2 displays the effective potentials for the 4D EGB black hole and the corresponding ATSW. For the black hole case, Fig.~2(a) illustrates three distinct types of photon trajectories in the inset at the bottom right corner. When $b < b_c$, the photon plunges into the black hole; when $b = b_c$, it moves on a circular orbit; and when $b > b_c$, the photon is gravitationally deflected and escapes. Figure~2(a) also shows that the height of the effective potential increases as the coupling constant $\alpha$ increases. The Novikov-Thorne accretion disk model in 4D EGB gravity was studied in Ref.~\cite{Heydari2021}.

For the ATSW case, Fig.~2(b), the effective potential $V_2(r_2)$ in spacetime $\mathcal{M}_2$ is rescaled by a factor of $Z^2$. For a given impact parameter $b_1$, three scenarios can be identified. In the special limit $\alpha \to 0$, the spacetime reduces to the Schwarzschild ATSW, corresponding to the black and green curves in Fig.~2(b). Thus, when $b_1 < Z b_{c_2}$, the photon starting from $\mathcal{M}_1$ passes through the throat into $\mathcal{M}_2$ and eventually reaches infinity on the $\mathcal{M}_2$ side. When $Z b_{c_2} < b_1 < b_{c_1}$, the photon enters $\mathcal{M}_2$, turns back at a turning point, returns to $\mathcal{M}_1$, and finally reaches infinity in $\mathcal{M}_1$. When $b_1 > b_{c_1}$, the photon remains entirely in $\mathcal{M}_1$, is reflected by the potential barrier, and returns to infinity in $\mathcal{M}_1$.

To study the observable appearance of the 4D EGB asymmetric thin-shell wormhole, we compute the photon trajectories and their deflection angles as they travel through the wormhole spacetime. Starting from Eq.~(10), the orbital equation for the photon can be expressed as follows:
\begin{equation}
\frac{1}{b_i^2} - \frac{f_i(r_i)}{r_i^2} = \frac{1}{r_i^4} \left( \frac{\mathrm{d}r_i}{\mathrm{d}\phi_i} \right)^2 .
\end{equation}

To simplify the calculation, we introduce the variable $x = 1/r$. The equation of motion for the photon then becomes:
\begin{equation}
G_i(x_i) = \frac{1}{b_i^2} - x_i^2 \left[1 + \frac{1 - \sqrt{1 + 8\alpha M_i x_i^3}}{2\alpha x_i^2}\right].
\end{equation}
\begin{figure}[htbp]
    \centering
    \begin{minipage}[b]{0.4\textwidth}
        \centering
        \includegraphics[width=\linewidth]{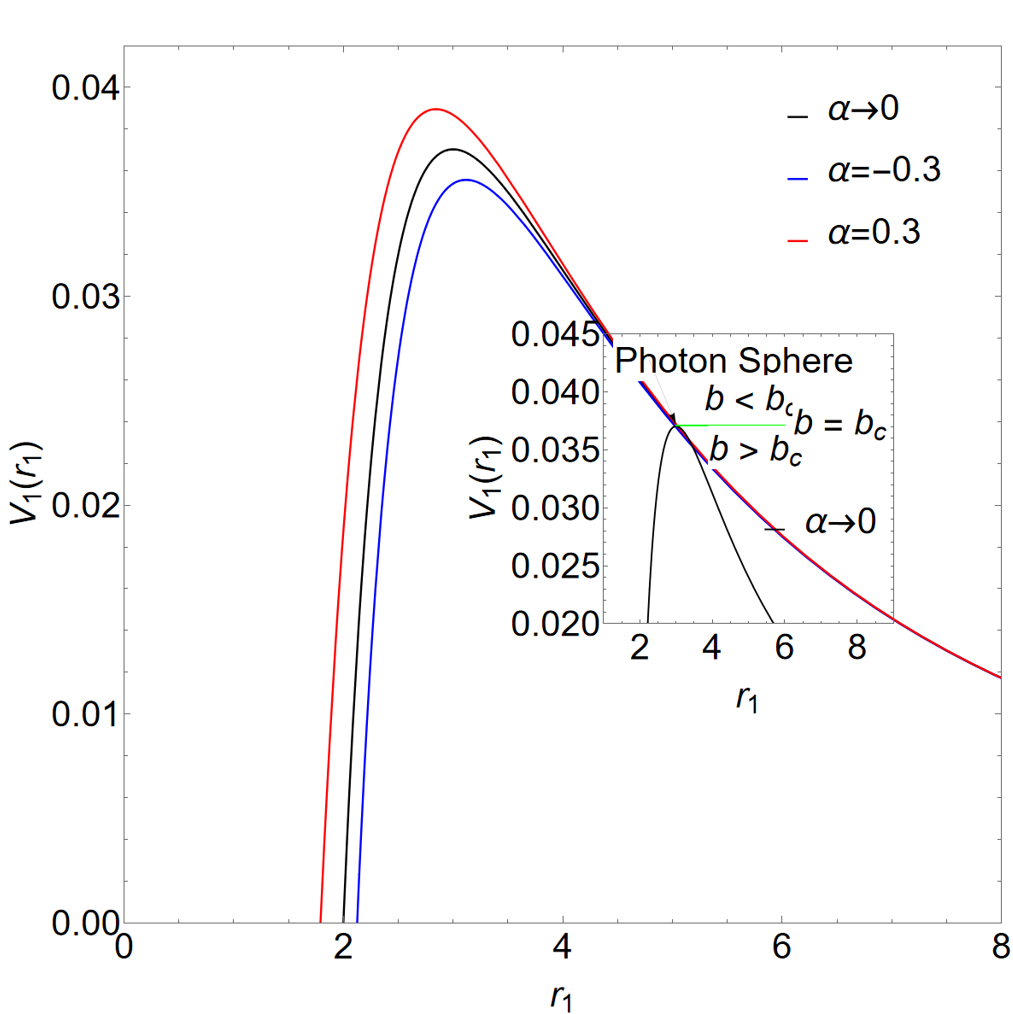}
        \small \textbf{(a)} $V(r)$ of the 4D EGB BH.
    \end{minipage}
    \hspace{1cm}   
    \begin{minipage}[b]{0.397\textwidth}
        \centering
        \includegraphics[width=\linewidth]{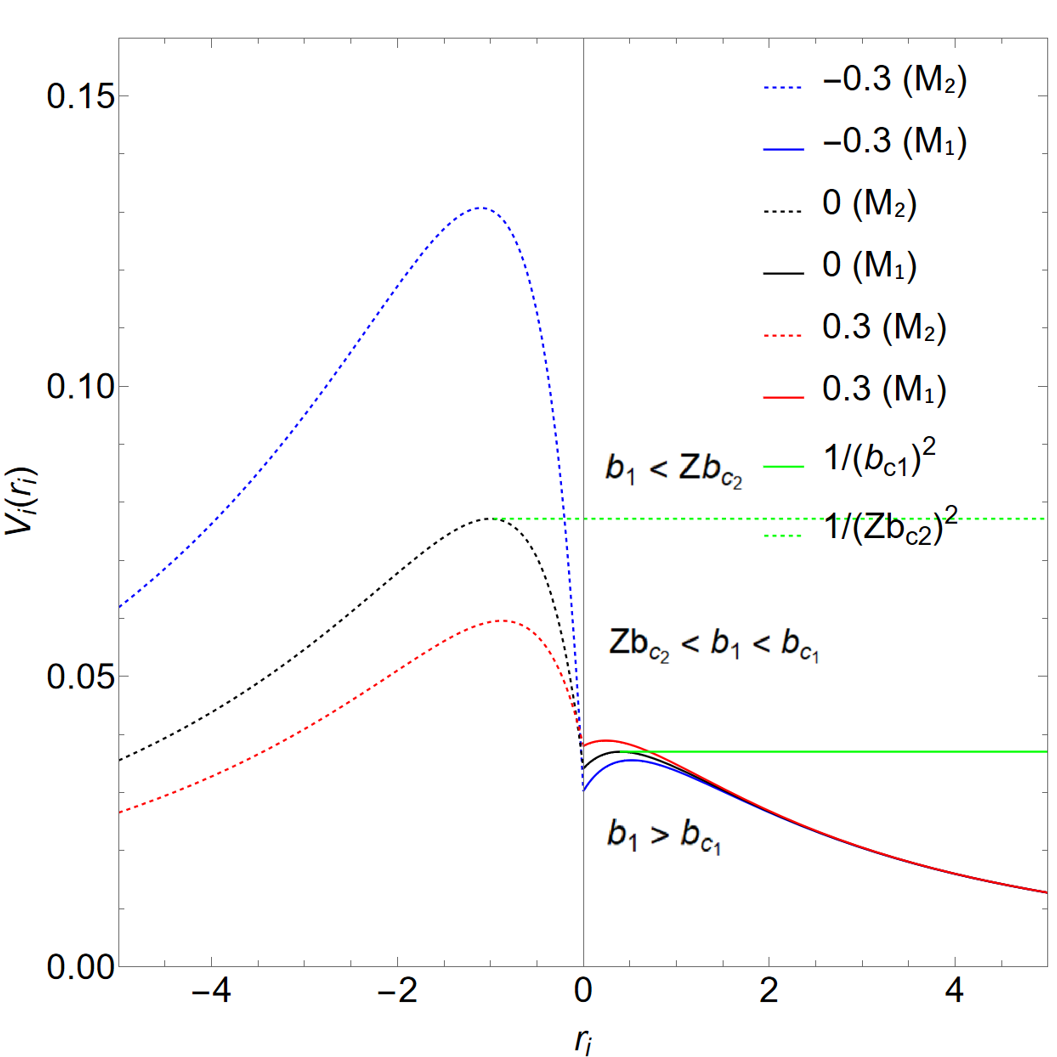}
        \small \textbf{(b)} $V(r)$ of the ATSW with 4D EGB profile.
    \end{minipage}
    \caption{(Color online) Effective potentials for the 4D EGB black hole (left panel) and the asymmetric thin-shell wormhole constructed from the 4D EGB metric (right panel). The Gauss--Bonnet coupling constant takes the values $\alpha = -0.3$, $\alpha \to 0$ (SS BH), and $\alpha = 0.3$. In the right panel, the solid and dashed curves represent the effective potentials in $\mathcal{M}_1$ and $\mathcal{M}_2$, respectively. In the limiting case $\alpha \to 0$ (the Schwarzschild ATSW), the critical impact parameters are $b_{c1} = 5.19615$ (green solid curve) and $Z b_{c2} = 3.6001$ (green dashed curve). The parameters are fixed at $M_1 = 1$, $M_2 = 1.2$, and $R = 2.6$.}
    \label{fig:Veff}
\end{figure}

When $b_1 < Z b_{c_2}$, the photon crosses the throat into $\mathcal{M}_2$ and eventually escapes to infinity on that side; therefore, its deflection angle need not be considered. For $b_1 > b_{c_1}$, the photon remains entirely in $\mathcal{M}_1$. The turning point in $\mathcal{M}_1$ is given by the smallest positive solution of $G_1(x_1) = 0$, denoted by $x_1^{\min}$. Using Eq.~(19), the total azimuthal change, i.e., the deflection angle, for a photon in $\mathcal{M}_1$ is obtained as
\begin{equation}
\phi_1(b_1) = 2\int_0^{x_1^{\min}} \frac{dx_1}{\sqrt{G_1(x_1)}}, \qquad b_1 > b_{c_1}.
\end{equation}
When the impact parameter lies in the interval $Z b_{c_2} < b_1 < b_{c_1}$, the photon passes through the throat, reaches a turning point in $\mathcal{M}_2$, and then returns to $\mathcal{M}_1$. For such a trajectory, the deflection angle accumulated in $\mathcal{M}_1$ is given by
\begin{equation}
\phi_1^*(b_1) = \int_0^{1/R} \frac{dx_1}{\sqrt{G_1(x_1)}}, \qquad b_1 < b_{c_1}. 
\end{equation}
In spacetime $\mathcal{M}_2$, the turning point is determined by the largest positive root of $G_2(x_2) = 0$, denoted by $x_2^{\max}$. The impact parameter $b_2$ is then obtained from Eq.~(17). Consequently, the deflection angle accumulated in $\mathcal{M}_2$ is expressed as
\begin{equation}
\phi_2(b_2) = 2\int_{x_2^{\max}}^{1/R} \frac{dx_2}{\sqrt{G_2(x_2)}}, \qquad b_2 > b_{c_2}. 
\end{equation}

Figure~3 shows the photon trajectories in the 4D EGB wormhole spacetime plotted using Eqs.~(20)--(22). For each value of $\alpha$ (from top to bottom: $-0.3$, $\alpha \to 0$, and $0.3$), the impact parameter is chosen within the range $Z b_{c_2} < b_1 < b_{c_1}$. For a photon originating from infinity in $\mathcal{M}_1$, its trajectory in $\mathcal{M}_2$ becomes longer as $b_1$ decreases. This trend holds for each $\alpha$, although it is illustrated only for representative values. Moreover, a comparison of the three rows in Fig.~3 shows that increasing the Gauss-Bonnet coupling constant $\alpha$ reduces $b_{c_1}$ while increasing $Z b_{c_2}$. These observations are consistent with the data listed in Table~I.

\section{Observational Appearance of The Asymmetric Thin-Shell Wormhole}\label{sec3}
Because of the wormhole's distinctive reflection properties, photons with a given impact parameter produce an observable image that differs significantly from that of a black hole. We assume that the 4D EGB ATSW is surrounded by a geometrically and optically thin accretion disk. Using two emission models, we compare the resulting wormhole images with those of a 4D EGB black hole.
\subsection{The trajectory of a photon}\label{sec3a}
We begin by defining the total orbit number of a photon, which measures the number of full cycles the photon completes around the black hole:
\begin{equation}
n = \frac{\phi}{2\pi}.
\end{equation}

Since the deflection angle depends on the impact parameter, the orbit number also depends on $b_1$. For a black hole, we place the observer at infinity at the north pole and the light source at infinity at the south pole, with the equatorial plane fixed. Based on the orbit number $n$, the photon paths are grouped into three categories~\cite{Gralla2019}:

Direct emission ($n < 0.75$): the photon crosses the equatorial plane exactly once;

Lensing ring ($0.75 < n < 1.25$): the photon crosses the equatorial plane twice;

Photon ring ($n > 1.25$): the photon crosses the equatorial plane at least three times.

For the optical appearance of the ATSW, the observer is still located at the north pole of $\mathcal{M}_1$. A photon with an impact parameter in the interval $Z b_{c_2} < b_1 < b_{c_1}$ originates in $\mathcal{M}_1$, passes through the throat, and enters $\mathcal{M}_2$. For such a wormhole spacetime, the orbit numbers are redefined as follows~\cite{Peng2021b}:
\begin{align}
n_1(b_1) &= \frac{\phi_1(b_1)}{2\pi},  \\
n_2(b_2) &= \frac{\phi_1(b_1) + \phi_2(b_1/Z)}{2\pi}, \\
n_3(b_1) &= \frac{2\phi_1(b_1) + \phi_2(b_1/Z)}{2\pi}. 
\end{align}
where $n_2$ and $n_3$ represent the additional orbit functions specific to the wormhole and give rise to the additional photon rings.
\begin{figure}[htbp]
\centering
\begin{minipage}[b]{0.3\textwidth}
    \centering
    \includegraphics[width=\linewidth]{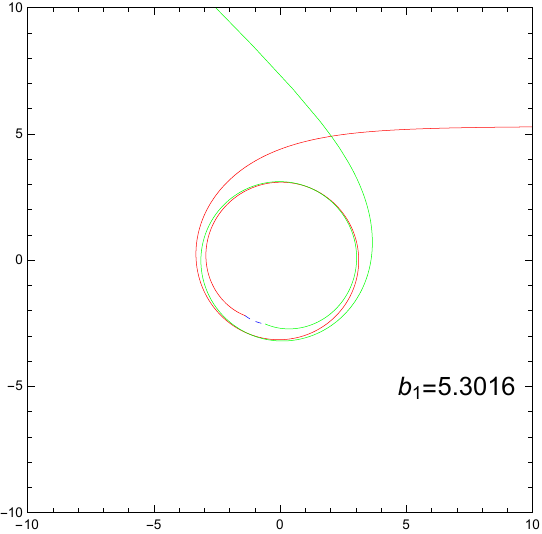}   
    \small (a) $\alpha = -0.3$, $b_{c1} = 5.30177$  
\end{minipage}
\hfill
\begin{minipage}[b]{0.3\textwidth}
    \centering
    \includegraphics[width=\linewidth]{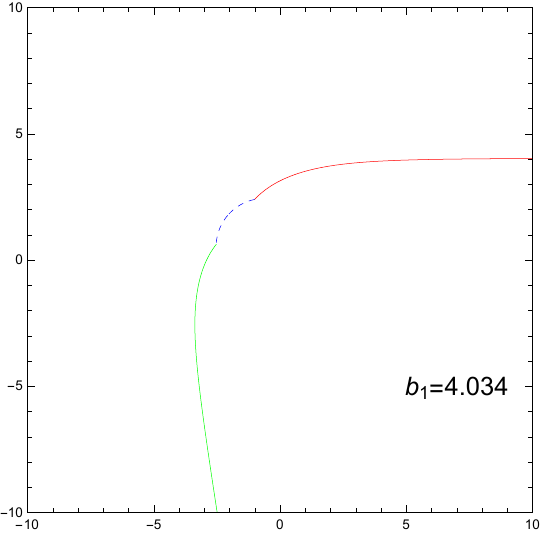}
    \small (b)  $\alpha = -0.3$
\end{minipage}
\hfill
\begin{minipage}[b]{0.3\textwidth}
    \centering
    \includegraphics[width=\linewidth]{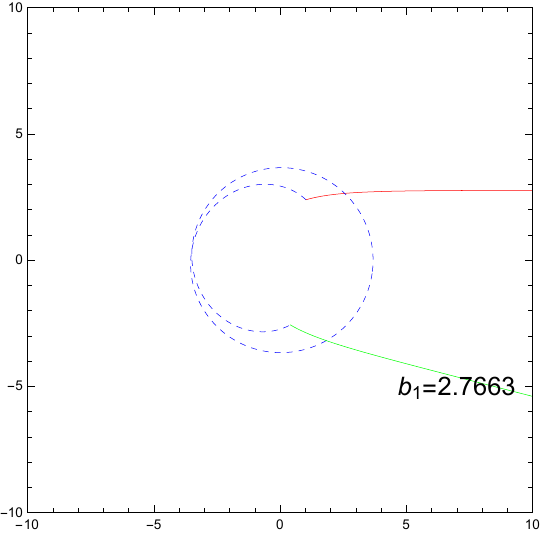}
    \small (c)  $\alpha = -0.3$, $Z b_{c_2} = 2.76623$
\end{minipage}

\vspace{5mm}  

\begin{minipage}[b]{0.3\textwidth}
    \centering
    \includegraphics[width=\linewidth]{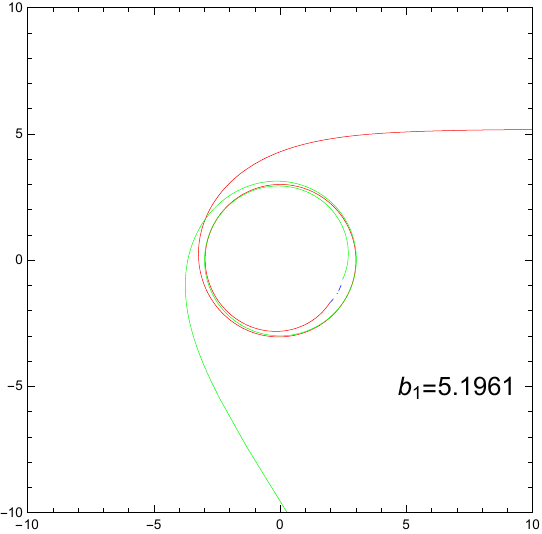}
    \small (d)  $\alpha \to 0$, $b_{c1} = 5.19615$
\end{minipage}
\hfill
\begin{minipage}[b]{0.3\textwidth}
    \centering
    \includegraphics[width=\linewidth]{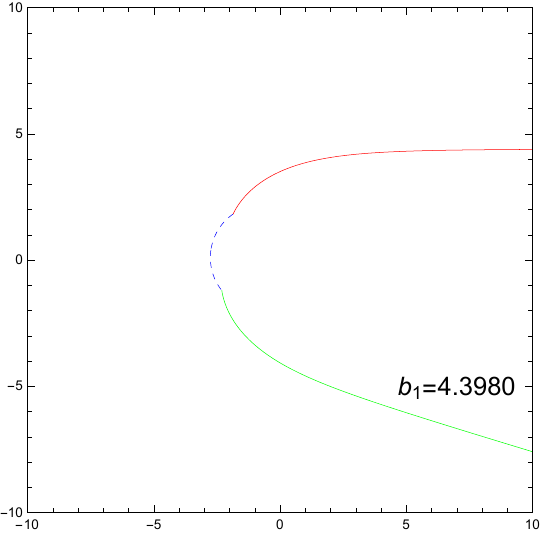}
    \small (e) $\alpha \to 0$
\end{minipage}
\hfill
\begin{minipage}[b]{0.3\textwidth}
    \centering
    \includegraphics[width=\linewidth]{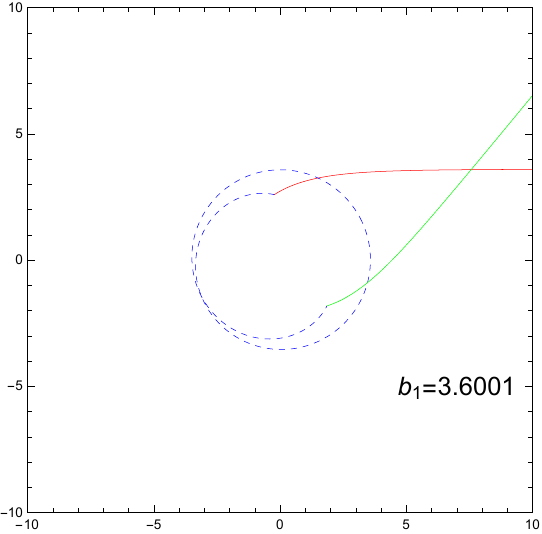}
    \small (f) $\alpha \to 0$, $Z b_{c_2} = 3.6$
\end{minipage}

\vspace{5mm}

\begin{minipage}[b]{0.3\textwidth}
    \centering
    \includegraphics[width=\linewidth]{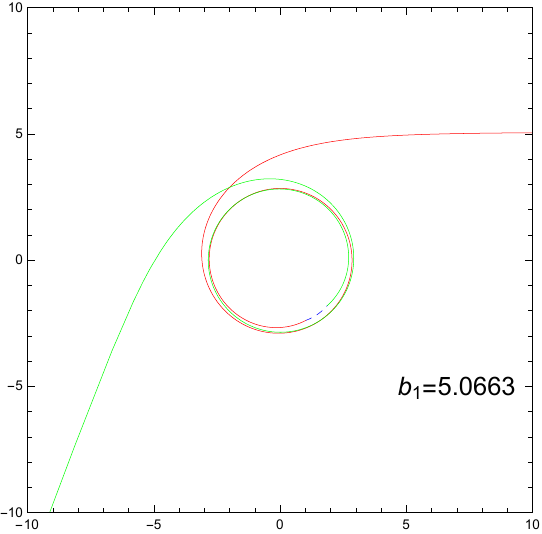}
    \small (g) $\alpha = 0.3$, $b_{c1} = 5.06641$
\end{minipage}
\hfill
\begin{minipage}[b]{0.3\textwidth}
    \centering
    \includegraphics[width=\linewidth]{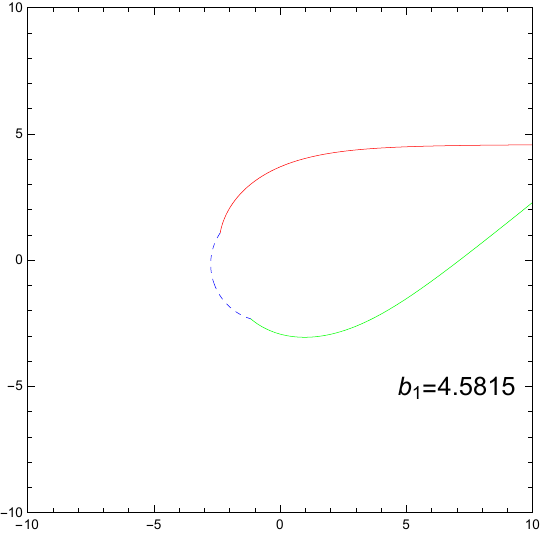}
    \small (h) $\alpha = 0.3$
\end{minipage}
\hfill
\begin{minipage}[b]{0.3\textwidth}
    \centering
    \includegraphics[width=\linewidth]{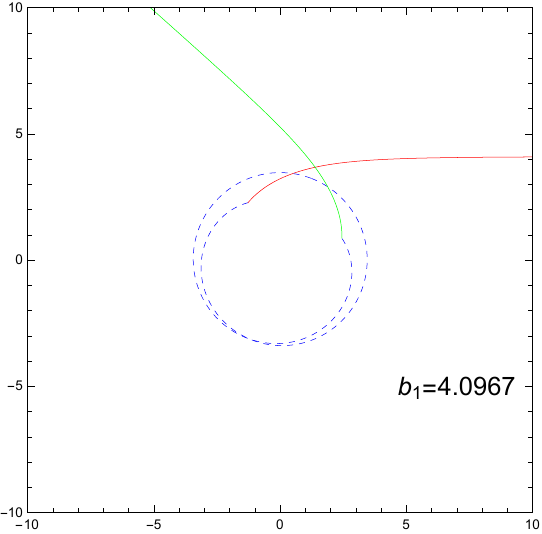}
    \small (i) $\alpha = 0.3$, $Z b_{c_2} = 4.09662$
\end{minipage}

\caption{(Color online) Representative light paths in the 4D EGB wormhole spacetime for impact parameters satisfying $Z b_{c_2} < b_1 < b_{c_1}$. The three rows correspond to different values of the Gauss--Bonnet parameter: $\alpha = -0.3$ (top), $\alpha \to 0$ (middle), and $\alpha = 0.3$ (bottom). In each panel, solid red curves denote photons approaching the throat from $\mathcal{M}_1$; dashed blue curves indicate motion inside $\mathcal{M}_2$; and solid green curves represent photons that re-emerge into $\mathcal{M}_1$ after reflection. All calculations use $M_1 = 1$, $M_2 = 1.2$, and $R = 2.6$.}  
\label{fig:traj}
\end{figure}

The behavior of $n_1$, $n_2$, and $n_3$ as functions of $b_1$ is shown in Fig.~4. Panel (a) shows that, for the wormhole, $n_1$ behaves similarly to the black-hole case: the photon never leaves $\mathcal{M}_1$. When the photon travels into $\mathcal{M}_2$ and subsequently returns, two additional orbit-number branches appear. Specifically, if $n_2 < 0.75$ and $n_3 > 0.75$ (solid and dashed curves in panel (b)), the photon intersects the far side of the accretion disk. If instead $n_2 < 1.25$ and $n_3 > 1.25$, the intersection occurs on the near side. Panel (b) further shows that, as $\alpha$ increases, the range of $b_1$ yielding these additional orbits decreases. Consequently, a larger $\alpha$ shifts the additional photon rings inward toward the central dark region of the wormhole shadow. This trend is fully consistent with the data in Tab.~I and the trajectories shown in Fig.~3.

\subsection{Observed intensity and transfer function}
We assume that the accretion flow forms a geometrically thin, optically thin disk in spacetime $\mathcal{M}_1$. A static observer at infinity is located at the north pole of $\mathcal{M}_1$, and the disk lies in the equatorial plane. In its rest frame, the disk emits isotropically. Because $\mathcal{M}_1$ is spherically symmetric, the emitted specific intensity depends only on the radial coordinate and is denoted by $I_{\nu}^{\mathrm{em}}(r)$, where $\nu$ is the emission frequency in the local static frame. For a distant observer receiving a specific intensity $I_{\nu'}^{\mathrm{obs}}$ at the redshifted frequency $\nu' = \sqrt{f(r)}\,\nu$, Liouville's theorem gives the invariant:
\begin{equation}
\frac{I_{\nu'}^{\mathrm{obs}}}{\nu'^3} = \frac{I_{\nu}^{\mathrm{em}}}{\nu^3}.
\end{equation}
Consequently, the observed intensity can be derived as follows:
\begin{equation}
I_{\nu'}^{\mathrm{obs}} = f^{3/2}(r) \, I_{\nu}^{\mathrm{em}}(r) .
\end{equation}
After integrating over frequency, the total observed intensity from a single intersection is
\begin{equation}
I^{\mathrm{obs}} = \int I_{\nu'}^{\mathrm{obs}} d\nu' = \int f^{2} I_{\nu}^{\mathrm{em}} d\nu = f^{2}(r) I^{\mathrm{em}}(r), 
\end{equation}
where $I^{\mathrm{em}}(r) = \int I_{\nu}^{\mathrm{em}} d\nu$ denotes the total emitted intensity. Summing over all intersections between the light ray and the disk, the total observed intensity is
\begin{equation}
I_{\mathrm{obs}}(b) = \sum_{n} I^{\mathrm{em}}(r)\, f^{2}(r) \Big|_{r = r_{n}(b_{1})} .
\end{equation}
Here, the function $r_n(b_1)$ gives the radial location at which the photon with impact parameter $b_1$ crosses the disk plane for the $n$-th time. The slope $dr_n/db_1$ is interpreted as the demagnification factor. Following the standard classification, the first transfer function ($n=1$) produces the ``direct emission'' (a redshifted version of the source profile), the second ($n=2$) generates the ``lensing ring'', and the third ($n=3$) yields the ``photon ring''.

The behavior of the transfer functions as functions of the impact parameter is shown in Fig.~5 for three values of the Gauss-Bonnet coupling constant: $\alpha = -0.3$, $\alpha \to 0$, and $\alpha=0.3$. Panel (a) shows the transfer functions for $\alpha = -0.3$. The first transfer function (black curve) gives a ``direct image'' with a small demagnification factor, reflecting the redshift of the source profile. The second transfer function (blue curve) produces a ``lensing ring'' characterized by a larger demagnification factor, corresponding to a reduced image of the far side of the disk. The third transfer function (red curve) exhibits the highest demagnification factor and yields the ``photon ring''---a highly compressed image of the near side of the disk.

In addition to the 4D EGB ATSW shown in the upper row, the lower row of Fig.~\ref{fig:three} presents the transfer functions of three comparison models with the same parameter values: (d) the Schwarzschild ATSW, (e) the Hayward ATSW with $g=0.3$, and (f) the Kalb-Ramond ATSW with $l=0$ and $Q=0.3$. The parameters are fixed at $M_1=1$, $M_2=1.2$, and $R=2.6$. All three panels show the transfer function $r/M_1$ as a function of the impact parameter $b_1/M_1$. The solid curves denote the usual transfer functions, which are also present for a black hole, whereas the dashed curves (blue and red) represent the new transfer functions arising from reflection at the wormhole throat. A comparison of panels (c), (e), and (f) reveals several differences. In terms of position, the extra transfer functions appear furthest to the right for the 4D EGB ATSW, at intermediate positions for the Hayward ATSW, and furthest to the left for the Kalb-Ramond ATSW. The slope of the third transfer function is steepest for the 4D EGB ATSW and gentlest for the Kalb-Ramond ATSW, which directly leads to differences in the luminosity of the third photon ring among the three models. The range of $b_1$ covered by the extra transfer functions is widest for the Kalb-Ramond ATSW and narrowest for the 4D EGB ATSW. Finally, the separation between the dashed and solid curves is most pronounced for the Kalb-Ramond ATSW and smallest for the 4D EGB ATSW.

Compared with a black hole, the ATSW possesses additional second transfer functions ($n=2$), plotted as blue dashed curves, which are referred to as the ``lensing band''. Near the critical curves $Z b_{c_2}$ and $b_{c_1}$, new third transfer functions ($n=3$, red dashed curves), known as the ``additional photon rings'', also appear.
\begin{figure}[H]
\centering
\begin{minipage}[b]{0.4\textwidth}
    \centering
    \includegraphics[width=\linewidth]{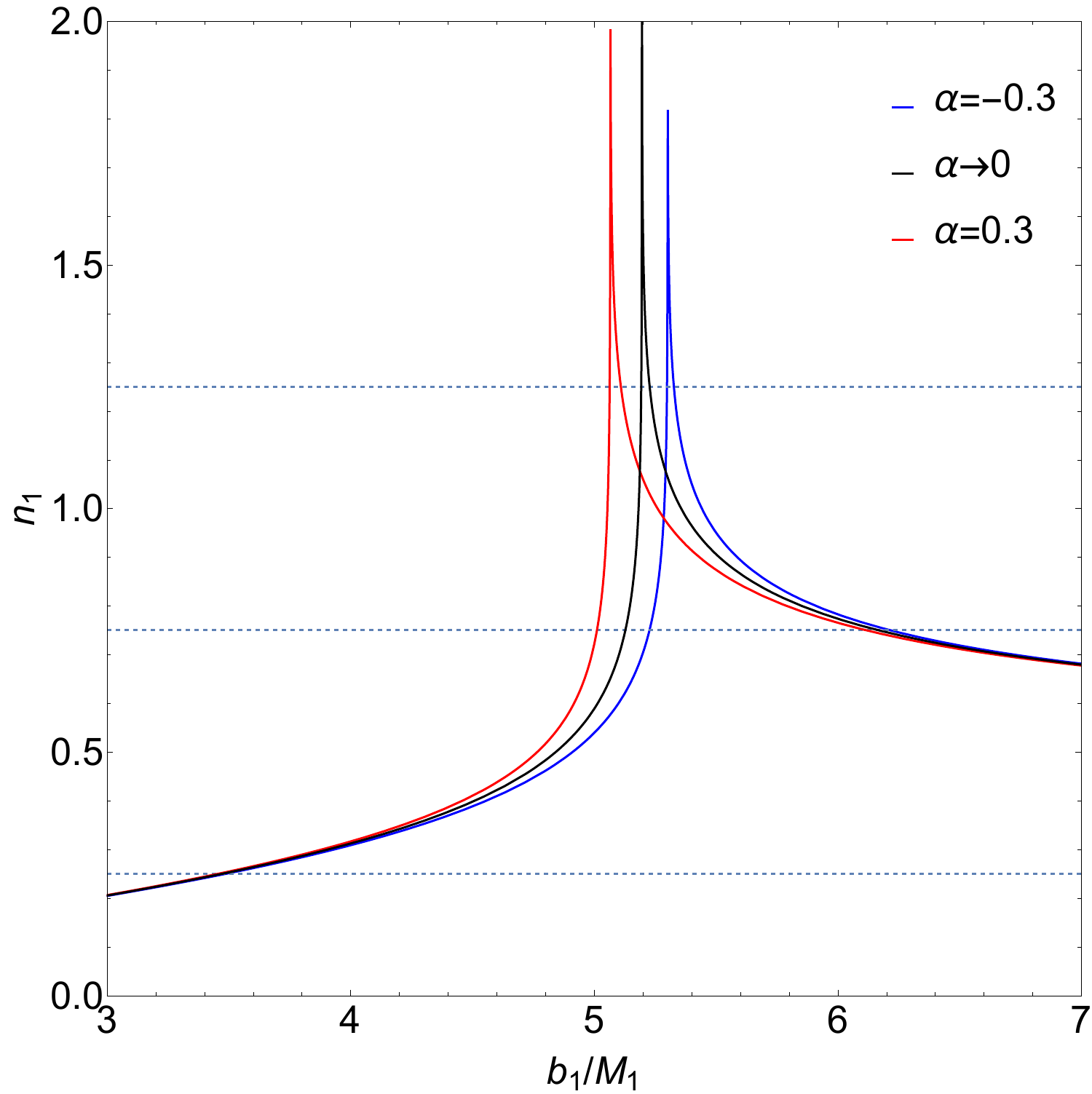}
    \small (a) The orbit number $n_1$.
\end{minipage}
\hspace{1cm}   
\begin{minipage}[b]{0.4\textwidth}
    \centering
    \includegraphics[width=\linewidth]{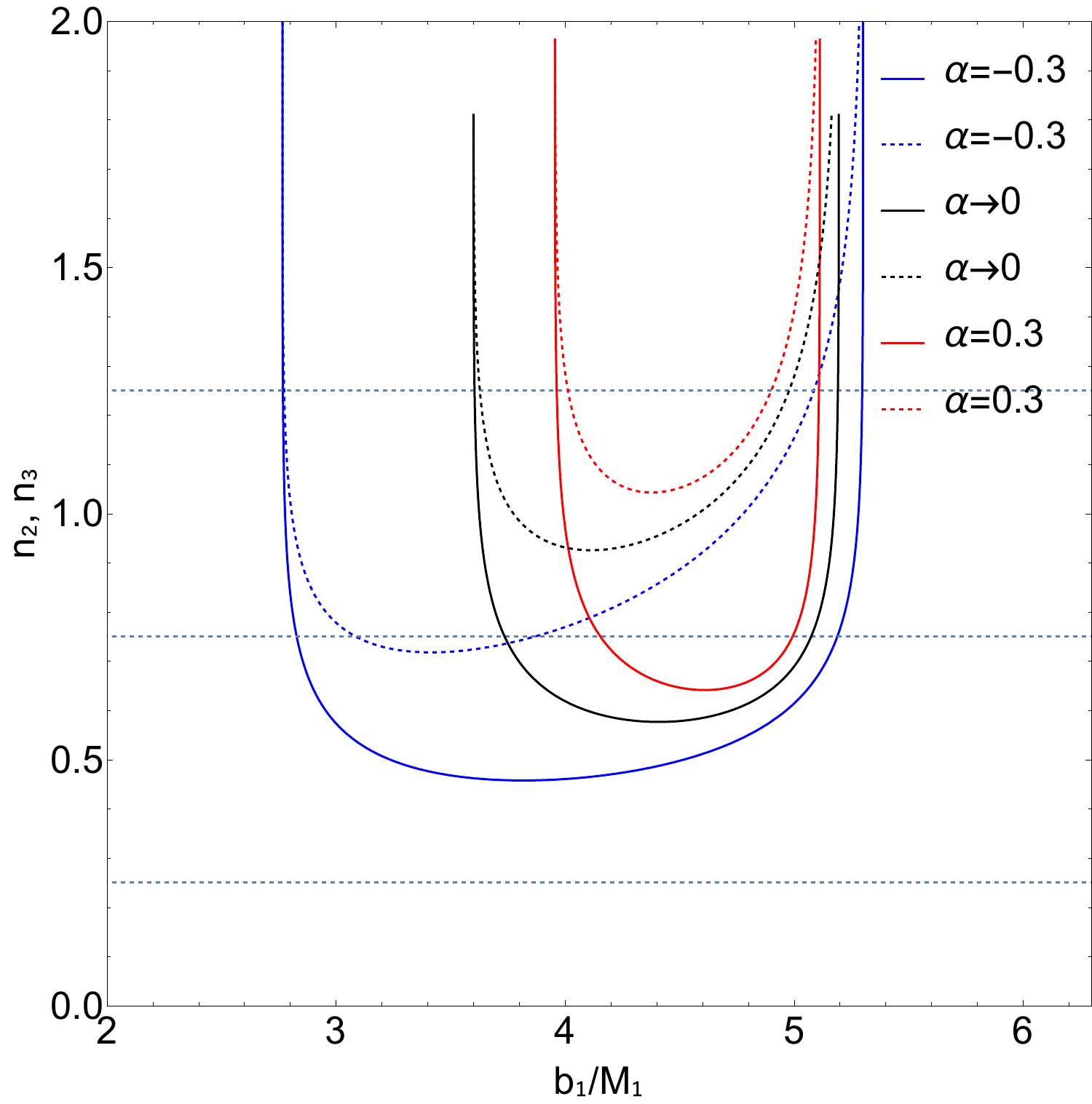}
    \small (b) The orbit number $n_2$, $n_3$.
\end{minipage}
\caption{(Color online) Photon orbit numbers around the ATSW with a 4D EGB profile. We consider different values of the Gauss--Bonnet coupling constant: $\alpha = -0.3$ (blue lines), $\alpha \to 0$ (black lines), and $\alpha = 0.3$ (red lines). The orbit number $n_1$ is shown in the left panel, whereas the orbit numbers $n_2$ (solid lines) and $n_3$ (dashed lines) are shown in the right panel. $M_1 = 1$, $M_2 = 1.2$, and $R = 2.6$.}
\label{fig:two}
\end{figure}

\begin{figure}[H]
\centering
\begin{minipage}[b]{0.3\textwidth}
    \centering
    \includegraphics[width=\linewidth]{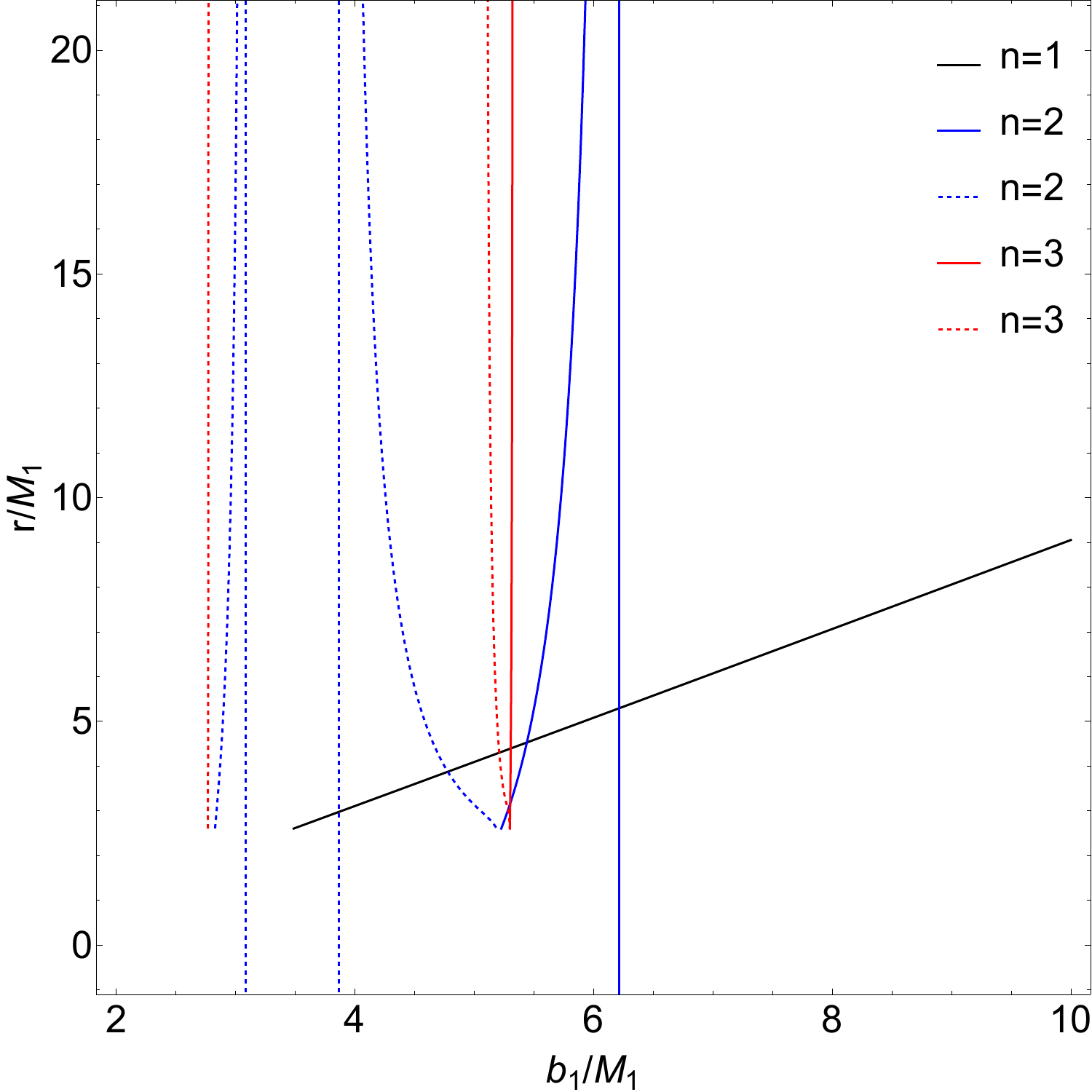}   
    \small (a) $\alpha = -0.3$   
\end{minipage}
\hfill
\begin{minipage}[b]{0.3\textwidth}
    \centering
    \includegraphics[width=\linewidth]{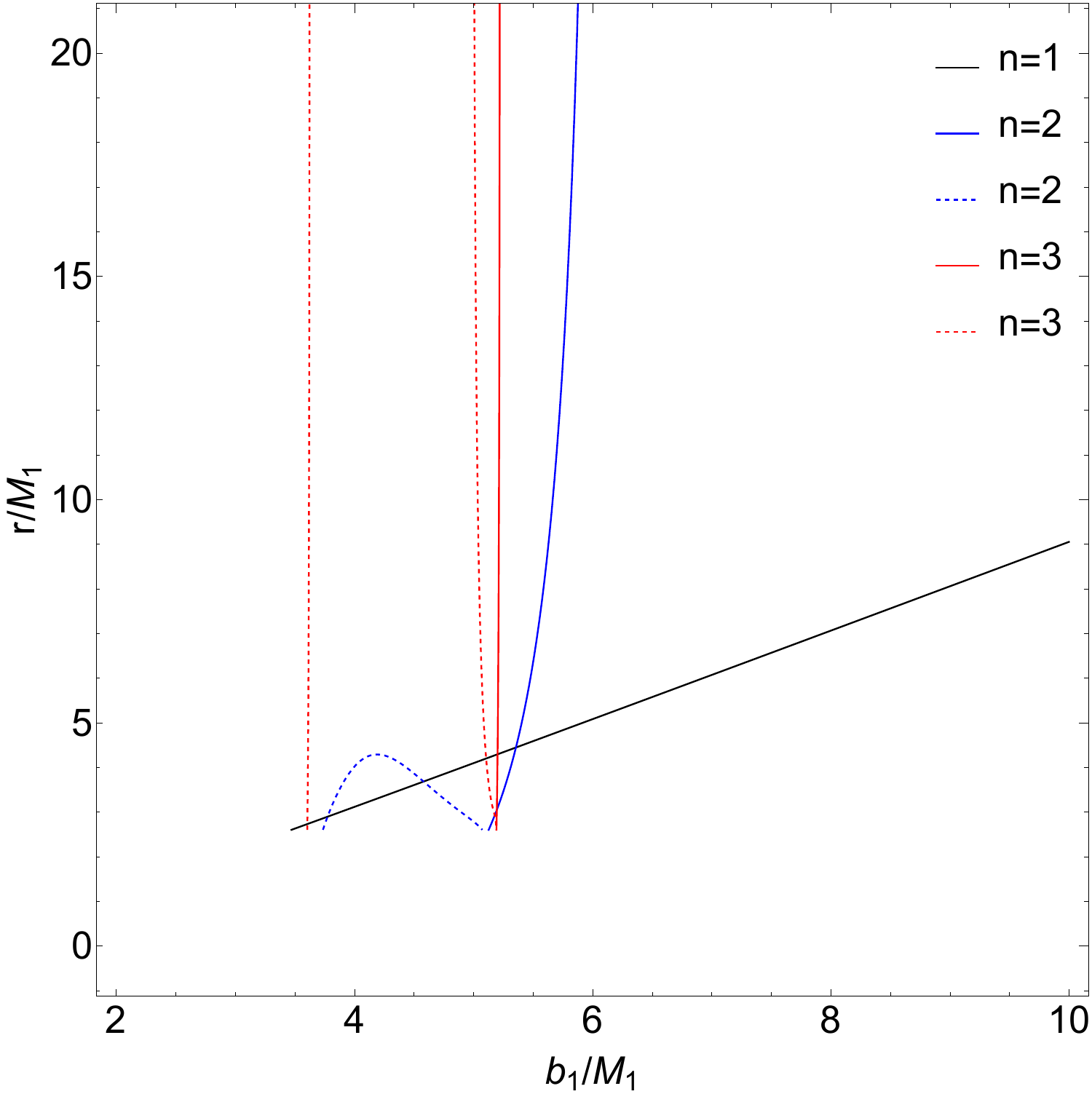}
    \small (b) $\alpha \to 0$
\end{minipage}
\hfill
\begin{minipage}[b]{0.3\textwidth}
    \centering
    \includegraphics[width=\linewidth]{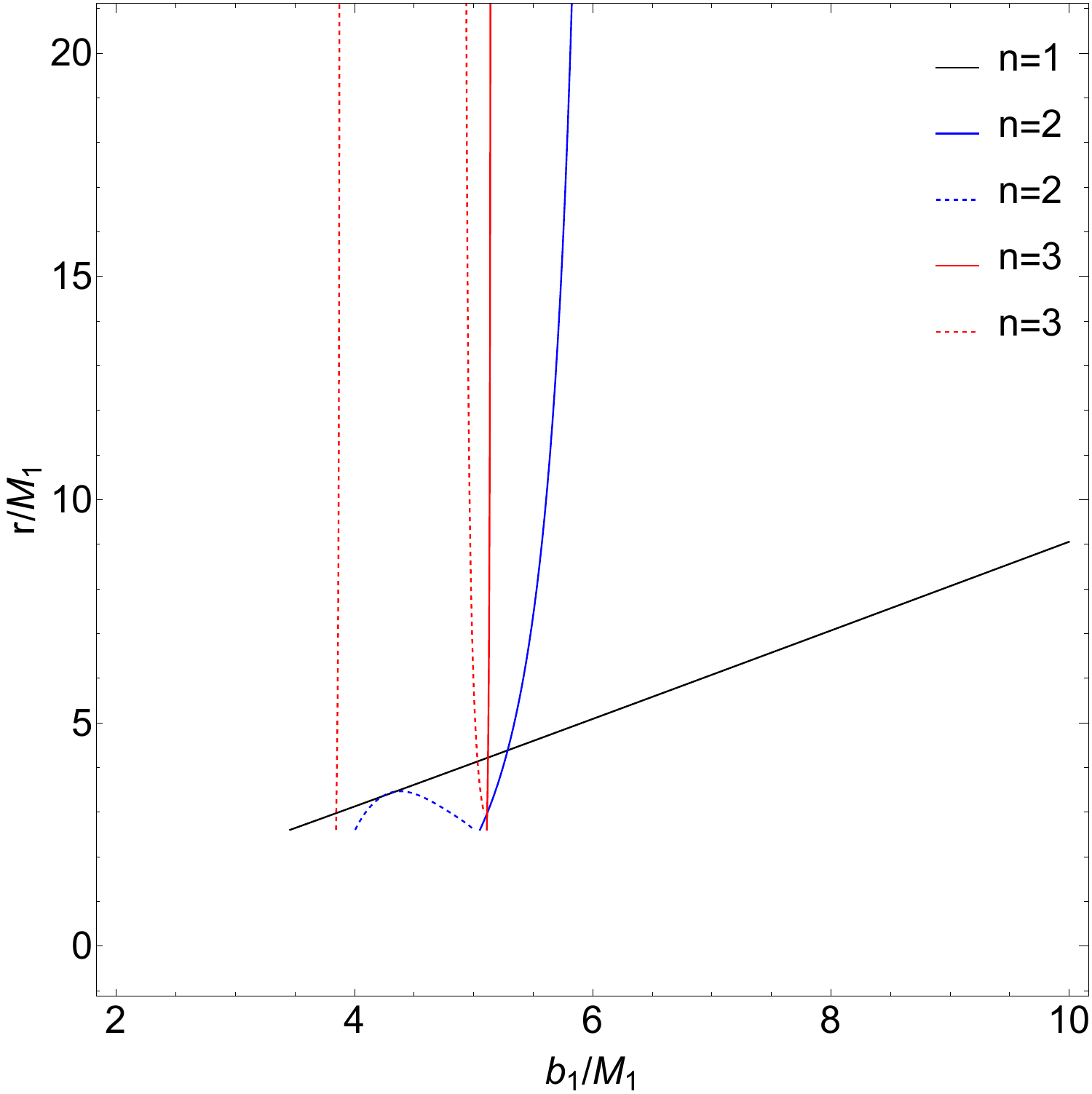}
    \small (c) $\alpha = 0.3$
\end{minipage}

\vspace{5mm}  

\begin{minipage}[b]{0.3\textwidth}
    \centering
    \includegraphics[width=\linewidth]{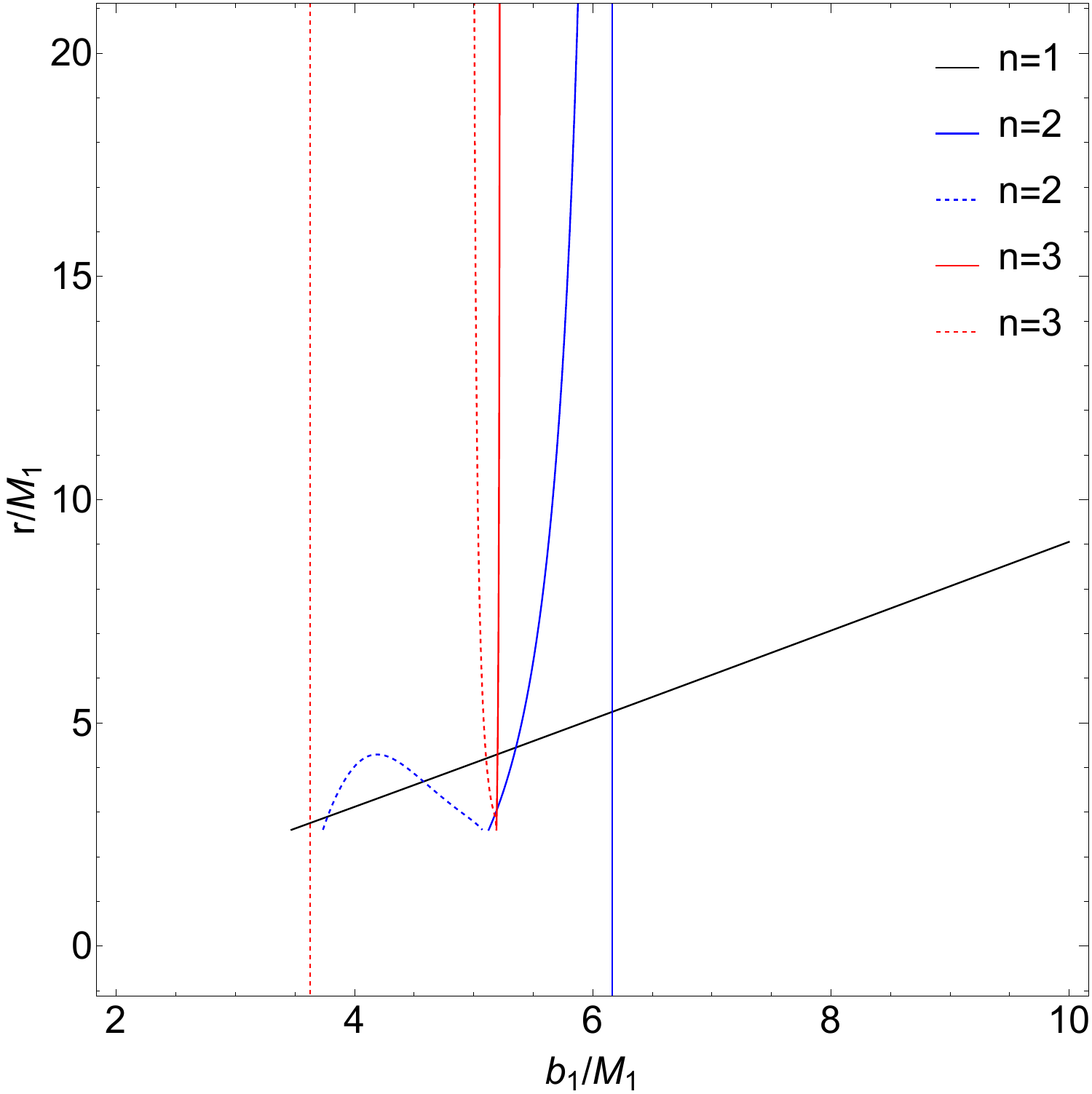}
    \small  (d) Schwarzschild ATSW
\end{minipage}
\hfill
\begin{minipage}[b]{0.3\textwidth}
    \centering
    \includegraphics[width=\linewidth]{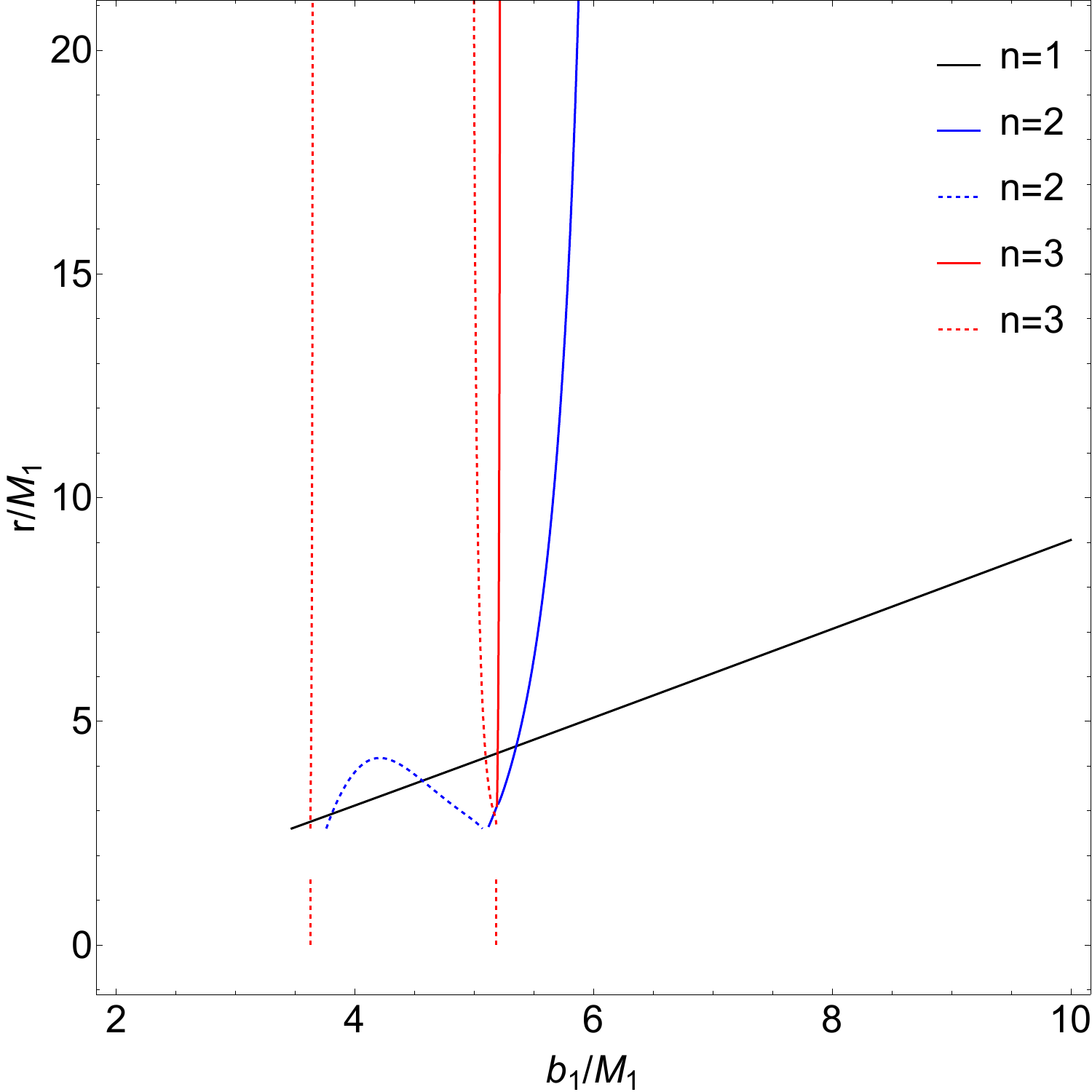}
    \small (e) Hayward ATSW, $g=0.3$
\end{minipage}
\hfill
\begin{minipage}[b]{0.3\textwidth}
    \centering
    \includegraphics[width=\linewidth]{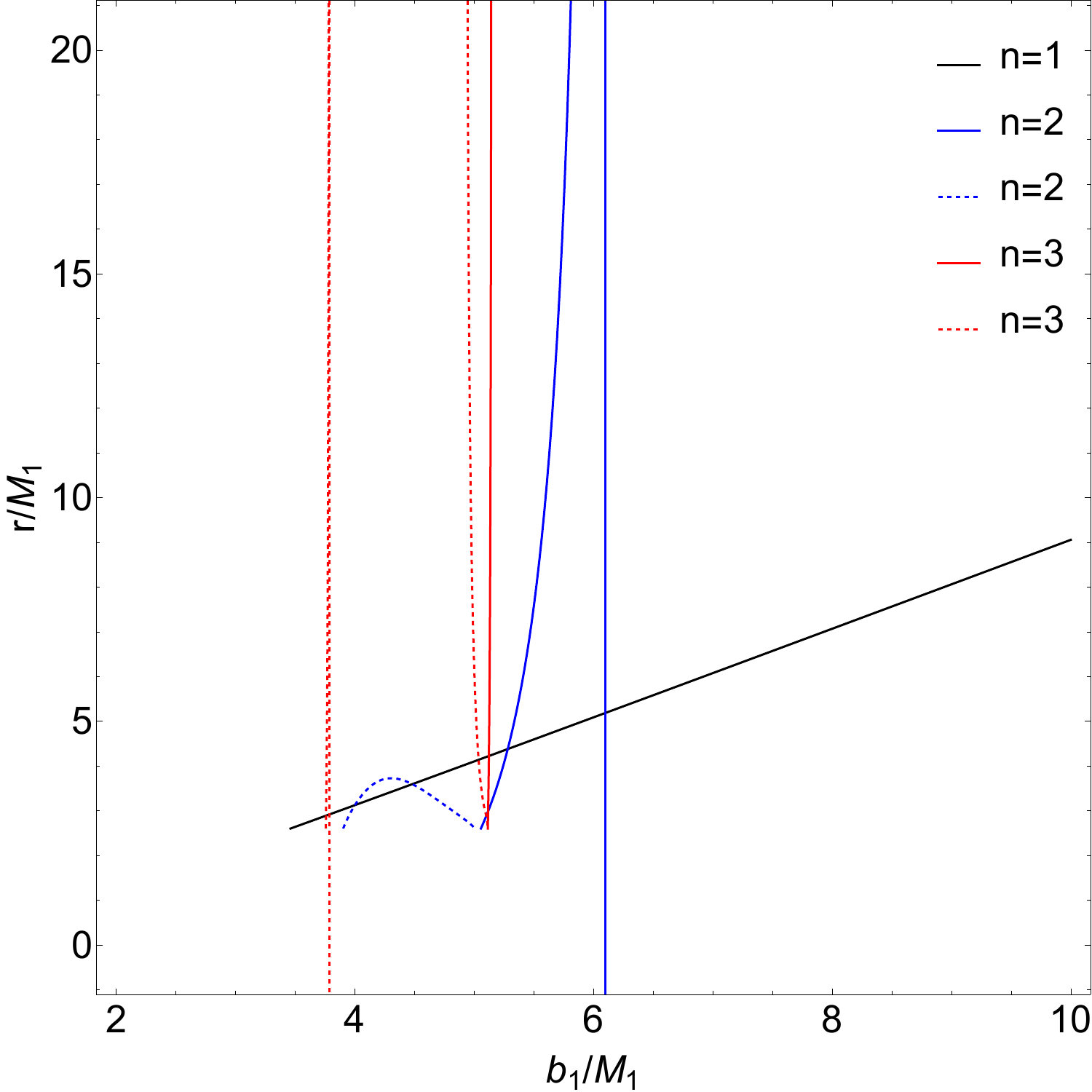}
    \small (f) Kalb–Ramond ATSW, $Q=0.3$
\end{minipage}

\caption{(Color online) Transfer functions for the 4D EGB asymmetric thin-shell wormhole. Three values of the Gauss-Bonnet coupling constant are considered: $\alpha=-0.3$, $\alpha\to 0$, and $\alpha=0.3$. The black, blue, and red curves denote the first, second, and third transfer functions, respectively. Solid curves represent the usual transfer functions, which are also present for a black hole, whereas the dashed blue and red curves denote new transfer functions arising from reflection at the wormhole throat. The parameters are $M_1=1$, $M_2=1.2$, and $R=2.6$. In addition, subfigures (d), (e), and (f) in the bottom row show the transfer functions for three comparison models with the same parameters: (d) the Schwarzschild ATSW, (e) the Hayward ATSW with $g=0.3$, and (f) the Kalb-Ramond ATSW with $l=0$ and $Q=0.3$.}
\label{fig:three}
\end{figure}

\subsection{Optical Appearance of Thin Accretion Disks under Two Emission Models in an Asymmetric Thin-Shell Wormhole}

In this section, two distinct emission models of the accretion disk are adopted to investigate the optical image characteristics of the ATSW. The observed image depends strongly on the specific form of the emission model. The emission from a thin accretion disk can be approximated by a Gaussian function. Setting $M_1 = 1$, the innermost stable circular orbit is computed numerically and denoted by $r_{\mathrm{isco}}$.
Emission model I is defined by

\begin{equation}
I_{\mathrm{I}}^{\mathrm{em}}(r) = \left\{ \begin{array}{ll}
0 & r < r_{\mathrm{isco}}, \\[6pt]
\left(\dfrac{1}{r - (r_{\mathrm{isco}} - 1)}\right)^2 & r \geq r_{\mathrm{isco}},
\end{array} \right. 
\end{equation}
where $r_{\mathrm{isco}}$ also denotes the inner edge of the disk, so that no emission originates from radii smaller than this value. This emission function is plotted in Fig.~6(a).

\begin{figure}[H]
\centering
\begin{minipage}[b]{0.4\textwidth}
    \centering
    \includegraphics[width=\linewidth]{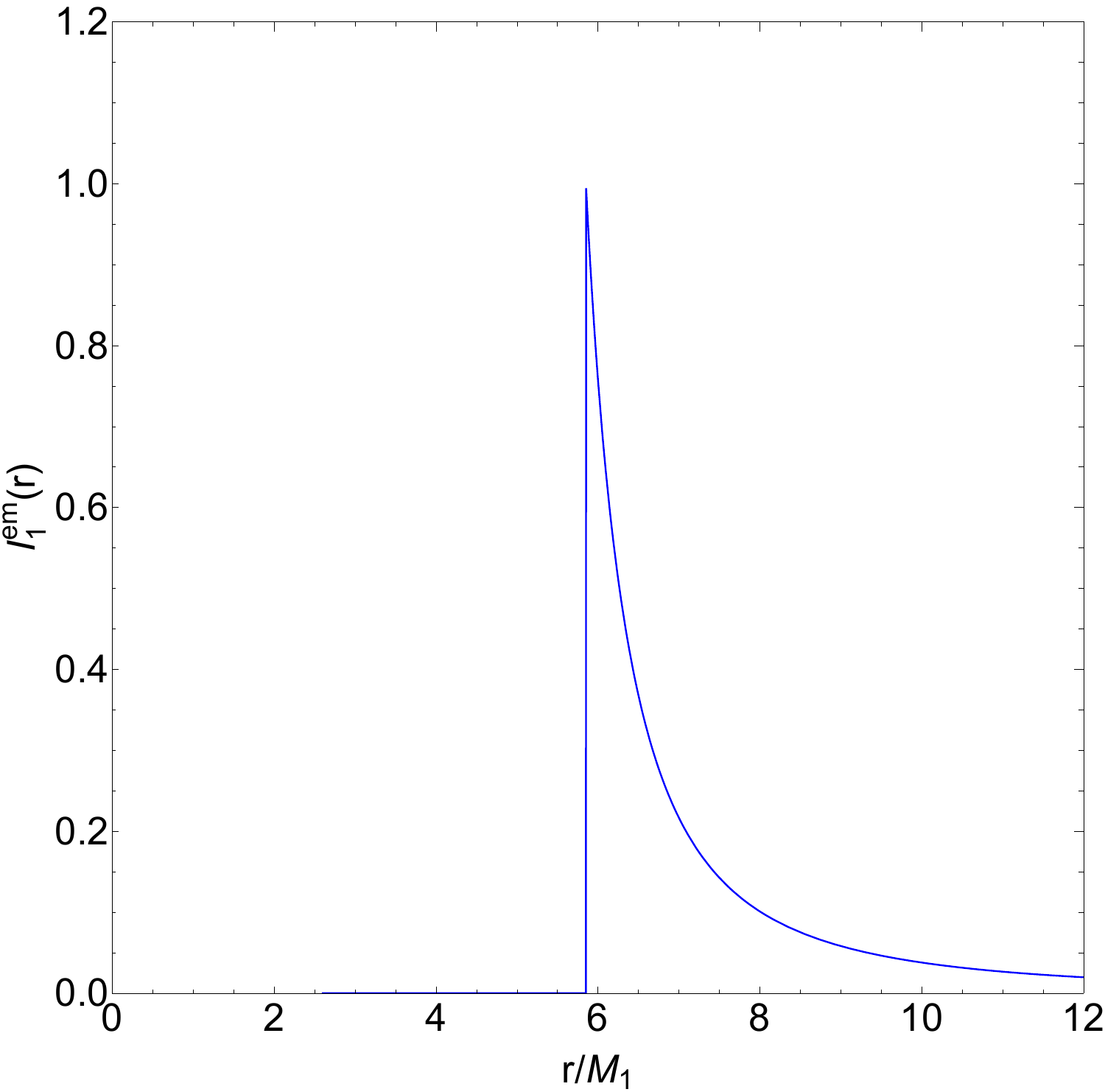}
    \small (a) Emission model I.
\end{minipage}
\hspace{1cm}  
\begin{minipage}[b]{0.4\textwidth}
    \centering
    \includegraphics[width=\linewidth]{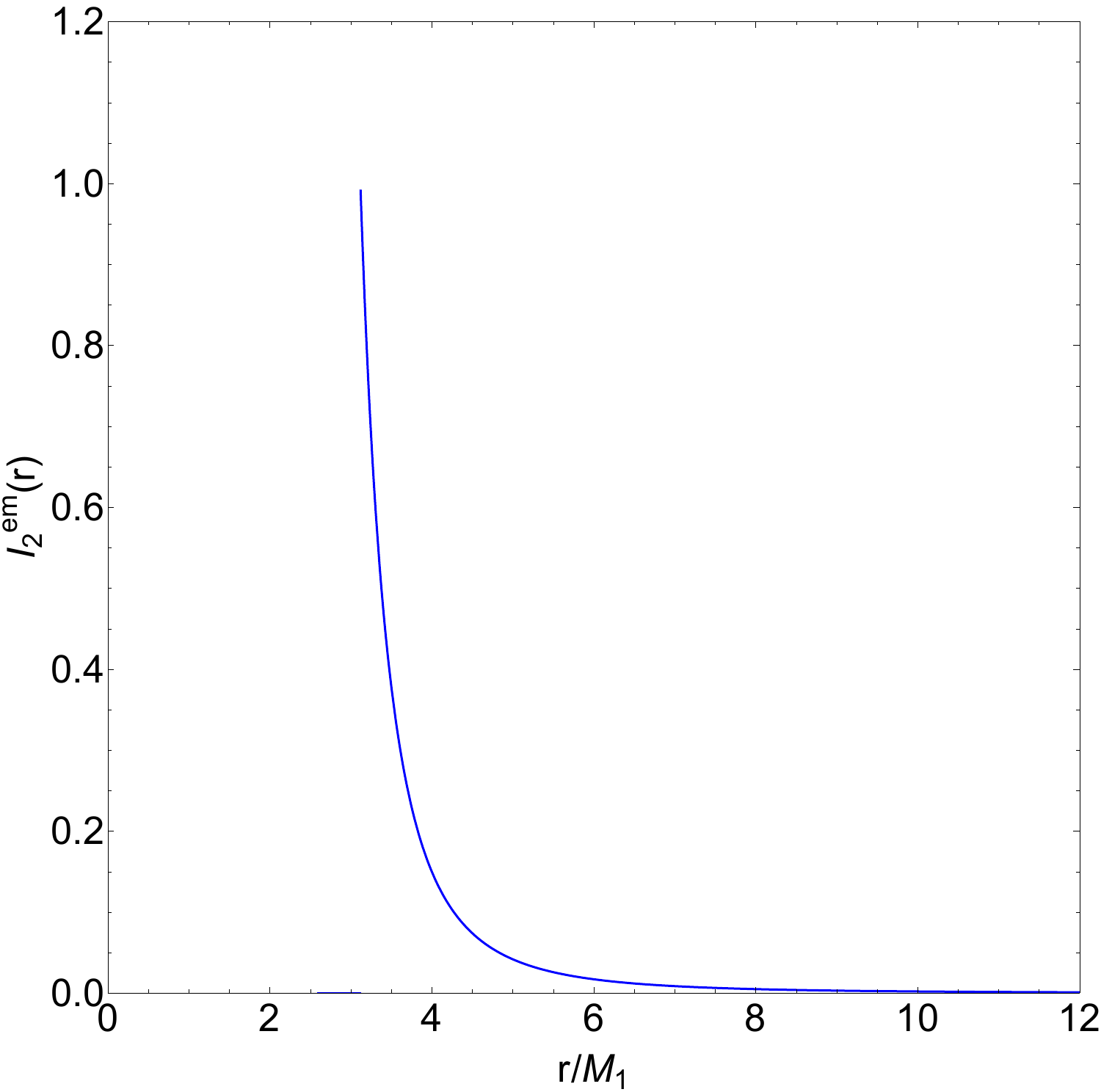}
    \small (b) Emission model II.
\end{minipage}
\caption{(color online) Two emission models for the accretion disk.}
\label{fig:five}
\end{figure}

Using model~I, we compute the observed intensity, density distribution, and local magnification for the ATSW; the results are shown in the upper panels of Fig.~7. For comparison, the lower panels of Fig.~7 show the corresponding images for a black hole with the same mass parameter and emission function.

\begin{figure}[htbp]
\centering
\begin{minipage}[b]{0.25\textwidth}
    \centering
    \includegraphics[width=\linewidth]{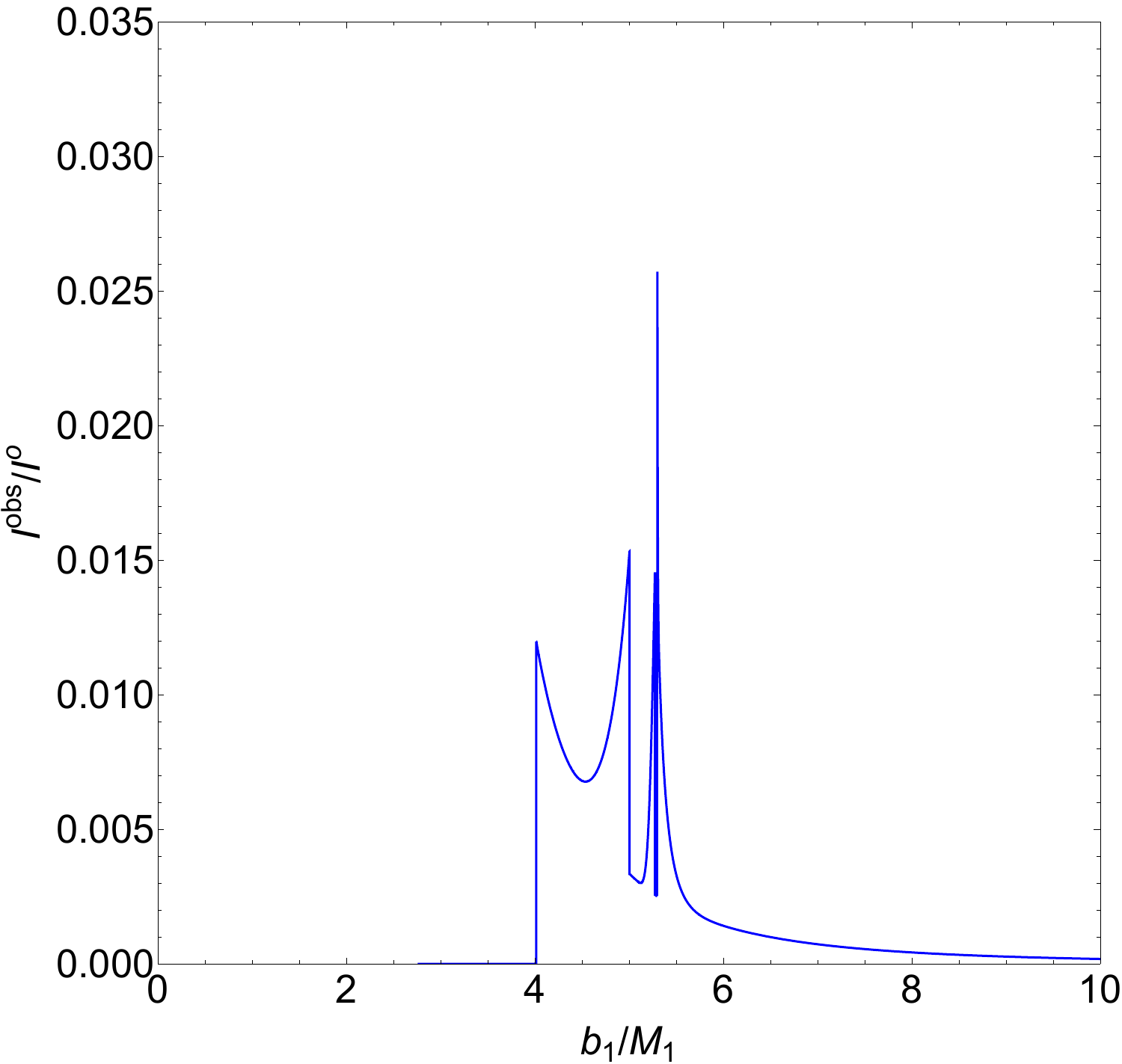}   
    \small (a)    
\end{minipage}
\hfill
\begin{minipage}[b]{0.25\textwidth}
    \centering
    \includegraphics[width=\linewidth]{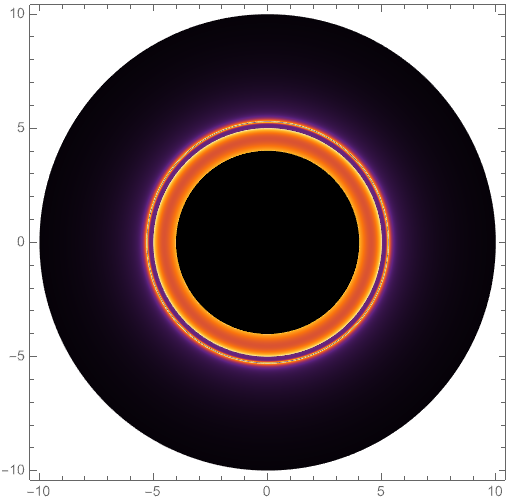}
    \small (b)
\end{minipage}
\hfill
\begin{minipage}[b]{0.25\textwidth}
    \centering
    \includegraphics[width=\linewidth]{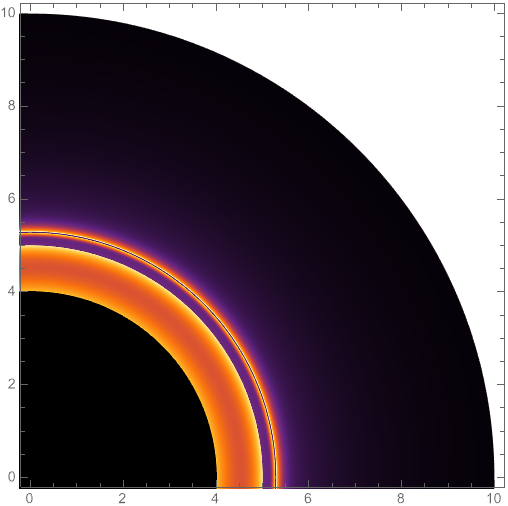}
    \small (c) 
\end{minipage}

\vspace{1mm}  

\begin{minipage}[b]{0.25\textwidth}
    \centering
    \includegraphics[width=\linewidth]{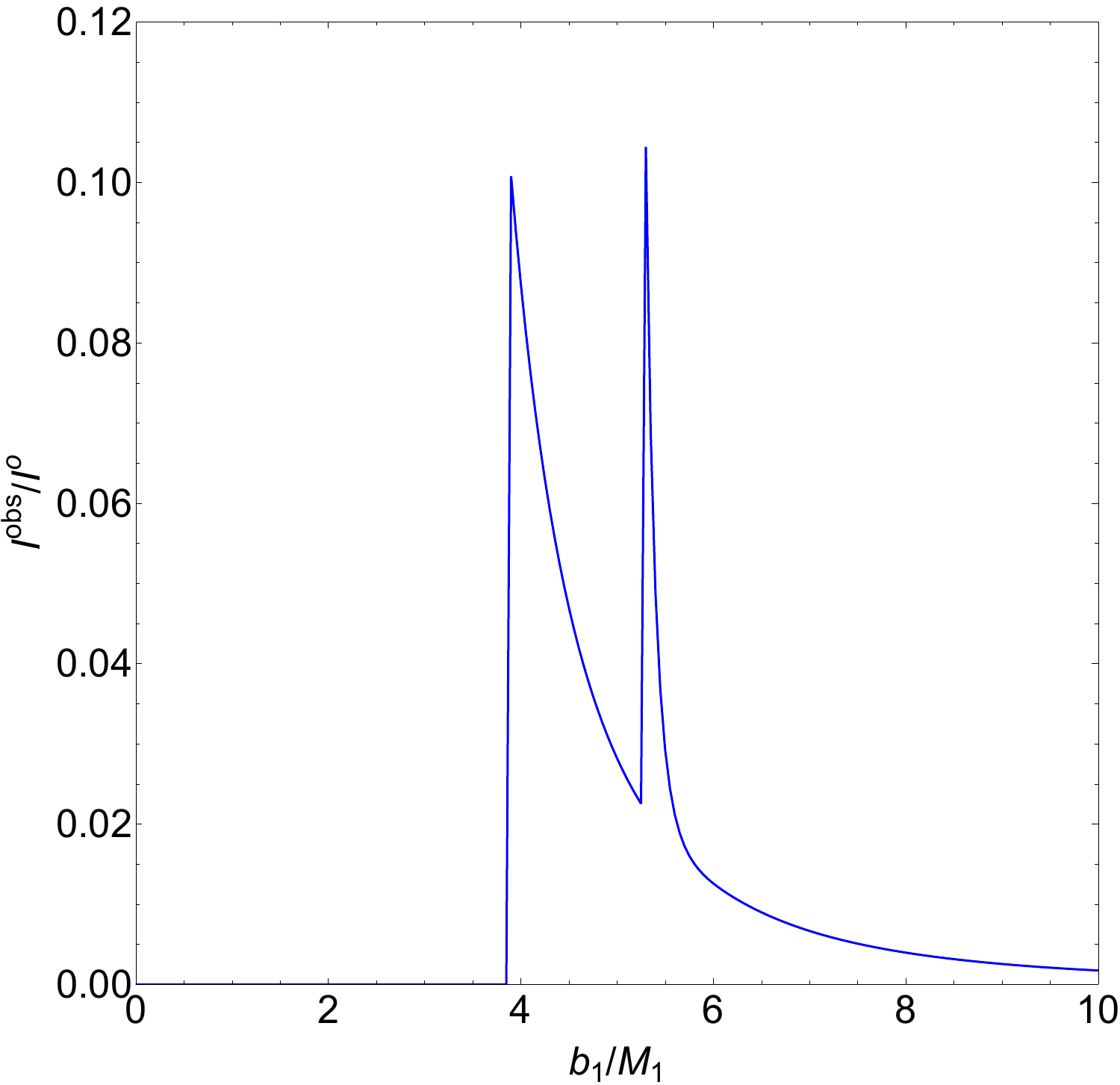}
    \small (d)
\end{minipage}
\hfill
\begin{minipage}[b]{0.25\textwidth}
    \centering
    \includegraphics[width=\linewidth]{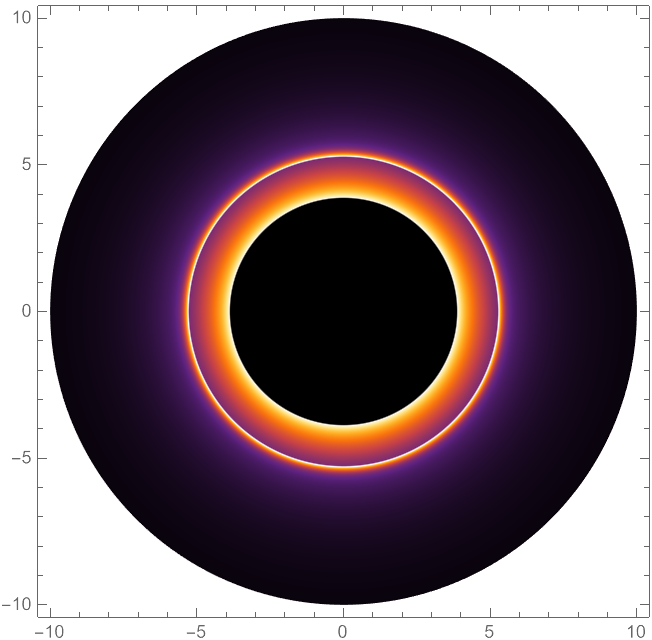}
    \small (e)
\end{minipage}
\hfill
\begin{minipage}[b]{0.25\textwidth}
    \centering
    \includegraphics[width=\linewidth]{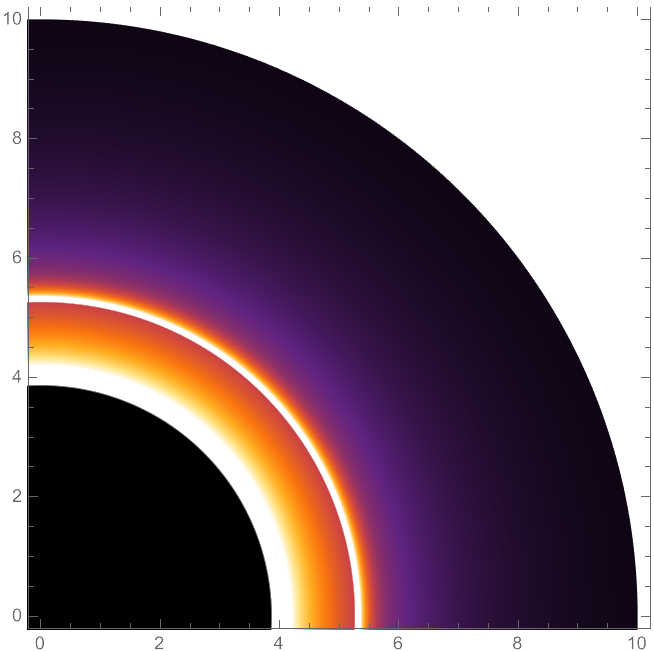}
    \small (f) 
\end{minipage}

\caption{(Color online) In emission model I: observed intensities (left panel), density plots (middle panel), and local density plots (right panel) for the 4D EGB ATSW (top row) and a 4D EGB black hole (bottom row). The Gauss--Bonnet coupling constant is $\alpha = -0.3$. The other parameters are $M_1 = 1$, $M_2 = 1.2$, and $R = 2.6$.}
\label{fig:six}
\end{figure}

Figs.~7(a) and 7(d) show that the direct emission, lensing band, and photon rings are spatially separated. For the ATSW (Fig.~7(a)), the direct emission appears near the critical curve at $b_1 \simeq 7.109M_1$ with an initial intensity of $0.454$, and then gradually decreases. The lensing band is confined between $b_1 \simeq 5.558M_1$ and $b_1 \simeq 5.995M_1$. The photon rings are located near $b_1 \simeq 4.457M_1$, $b_1 \simeq 5.169M_1$, and $b_1 \simeq 5.307M_1$. Compared with the black hole case (Figs.~7(a) and 7(d)), the ATSW in Fig.~7(a) exhibits two additional photon rings near $b_1 \simeq 4.457M_1$ and $b_1 \simeq 5.169M_1$. The density plot and its local magnification (Figs.~7(b) and 7(c)) show that the direct emission lies at the periphery of the black disk, whereas the narrow lensing band is enclosed within the black disk. In contrast to the black hole case (Figs.~7(e) and 7(f)), the ATSW exhibits two additional photon rings near the center of the black disk.

\begin{figure}[htbp]
\centering
\begin{minipage}[b]{0.25\textwidth}
    \centering
    \includegraphics[width=\linewidth]{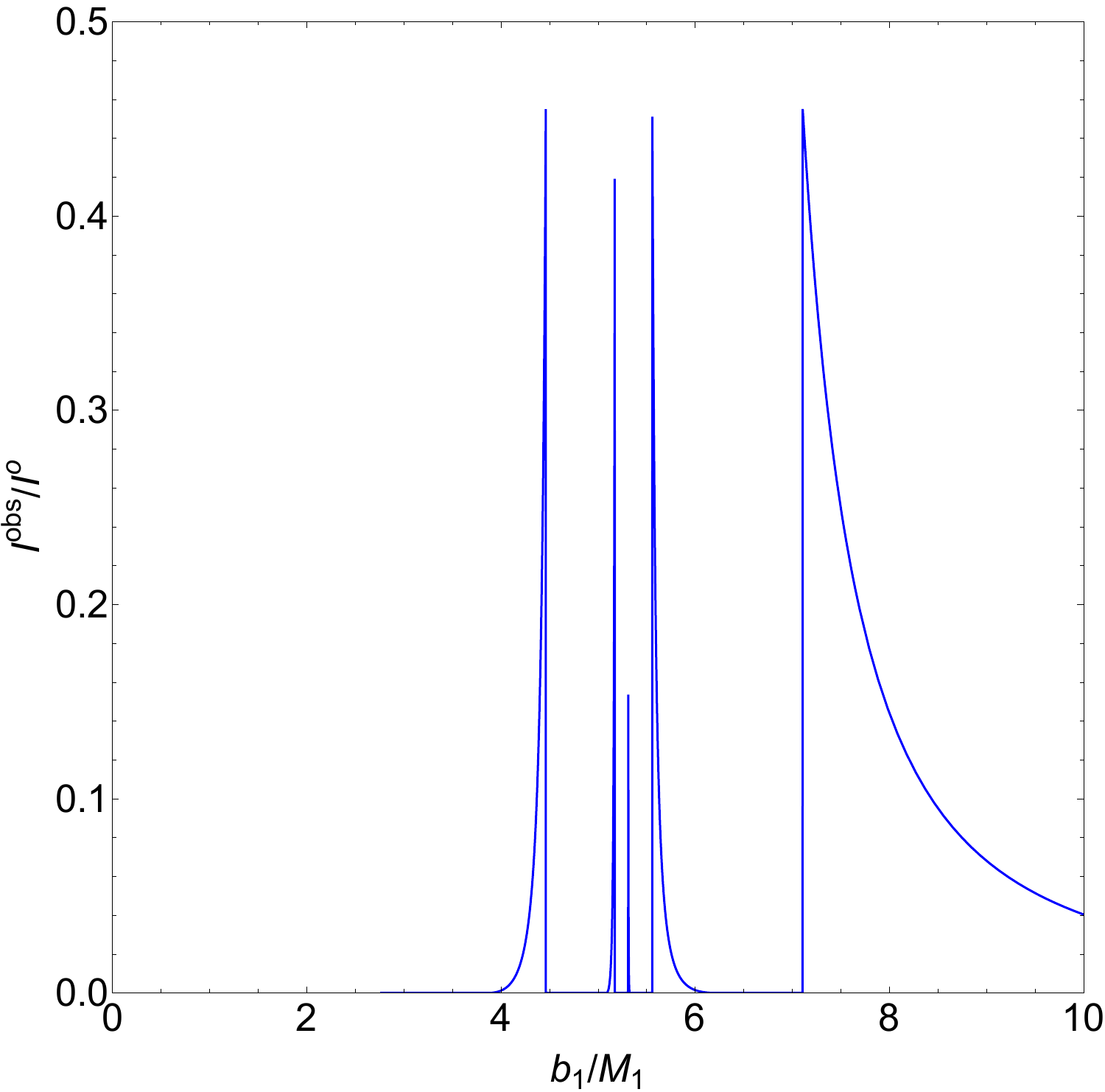}   
    \small (a)   
\end{minipage}
\hfill
\begin{minipage}[b]{0.25\textwidth}
    \centering
    \includegraphics[width=\linewidth]{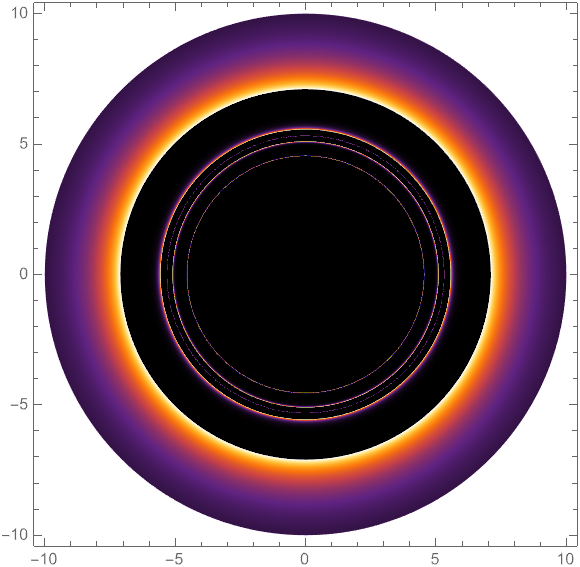}
    \small (b) 
\end{minipage}
\hfill
\begin{minipage}[b]{0.25\textwidth}
    \centering
    \includegraphics[width=\linewidth]{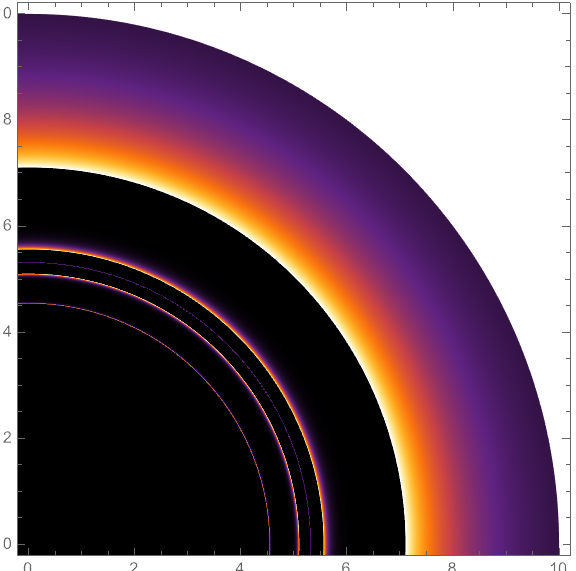}
    \small (c) 
\end{minipage}

\vspace{1mm}  

\begin{minipage}[b]{0.25\textwidth}
    \centering
    \includegraphics[width=\linewidth]{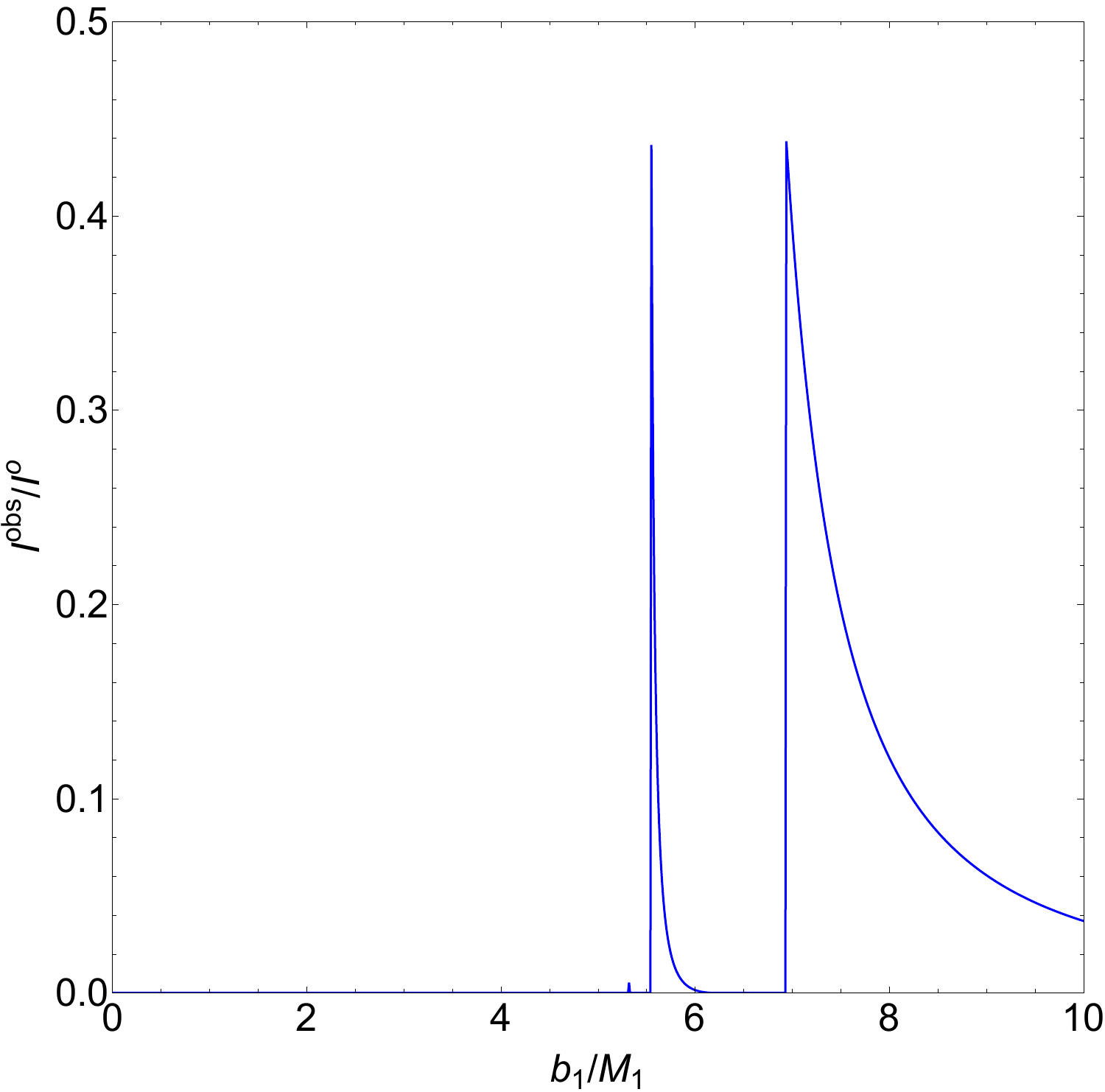}
    \small (d) 
\end{minipage}
\hfill
\begin{minipage}[b]{0.25\textwidth}
    \centering
    \includegraphics[width=\linewidth]{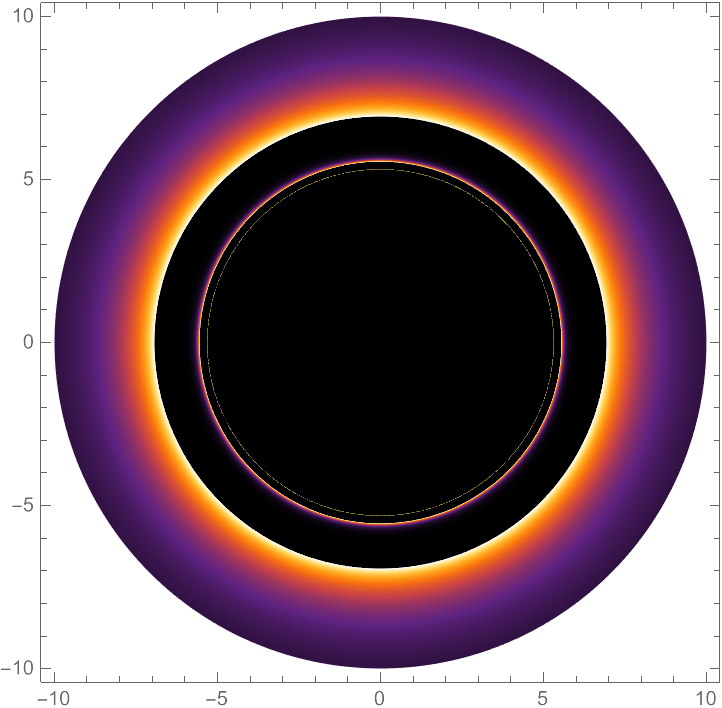}
    \small (e) 
\end{minipage}
\hfill
\begin{minipage}[b]{0.25\textwidth}
    \centering
    \includegraphics[width=\linewidth]{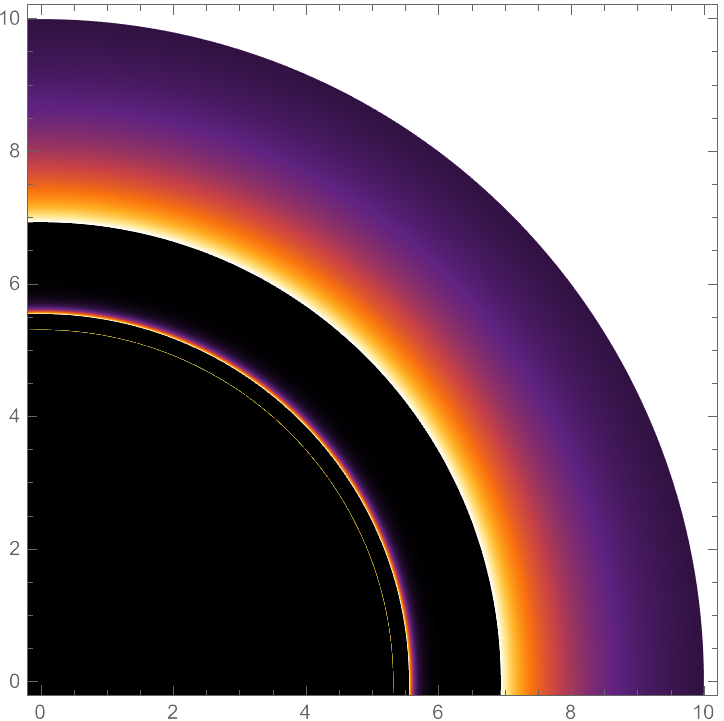}
    \small (f) 
\end{minipage}

\caption{(color online) For emission model II, the observed intensities (left panels), density plots (middle panels), and local density plots (right panels) are shown for the 4D EGB ATSW (top row) and a 4D EGB black hole (bottom row). The Gauss--Bonnet coupling constant is set to $\alpha = -0.3$. The remaining parameters are $M_1 = 1$, $M_2 = 1.2$, and $R = 2.6$.}
\label{fig:seven}
\end{figure}

Emission model II is defined by
\begin{equation}
I_{\mathrm{II}}^{\mathrm{em}}(r) = 
\begin{cases}
0, & r < r_{ph}, \\[6pt]
\displaystyle \frac{1}{\bigl(r - (r_{ph} - 1)\bigr)^3}, & r \ge r_{ph},
\end{cases}
\end{equation}
where $r_{ph}$ is the radius of the photon sphere of the 4D EGB black hole. This model decays more rapidly than model~I, as shown in Fig.~6(b). Using model~II, we compute the observed intensity, density map, and local magnification for the ATSW (upper row of Fig.~8) and the black hole (lower row of Fig.~8).

Figures~8(a) and 8(d) show that the direct emission, lensing band, and photon rings overlap in both the wormhole and black hole cases. For the ATSW, the direct emission is located near the critical curve $b_1 \simeq 4.015M_1$. A bright multilayered ring structure appears because the photon rings are embedded within the lensing band. Similar bright ring structures have been observed in images of naked singularities in EGB gravity~\cite{Gyulchev2021}. The corresponding black hole image is shown in the bottom row of Fig.~8. Figure~8 shows an additional lensing band between the critical curves $Z b_{c_2} \simeq 2.766M_1$ and $b_{c_1} \simeq 5.30178M_1$ for emission model~II, indicating that the new second transfer function contributes to the observed intensity of the ATSW.

\begin{figure}[H]
\centering
\begin{minipage}[b]{0.25\textwidth}
    \centering
    \includegraphics[width=\linewidth]{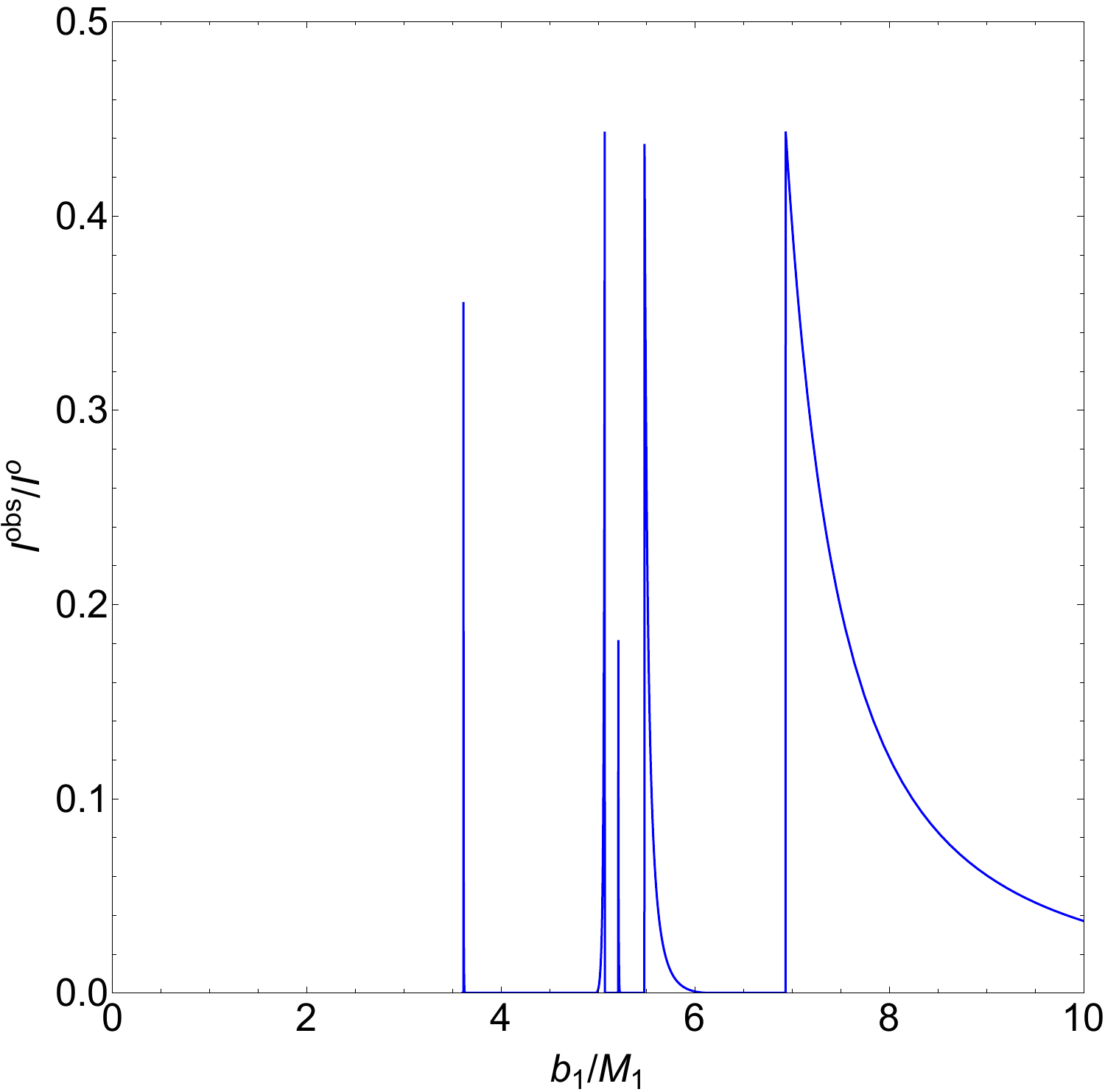}   
    \small (a)   
\end{minipage}
\hfill
\begin{minipage}[b]{0.25\textwidth}
    \centering
    \includegraphics[width=\linewidth]{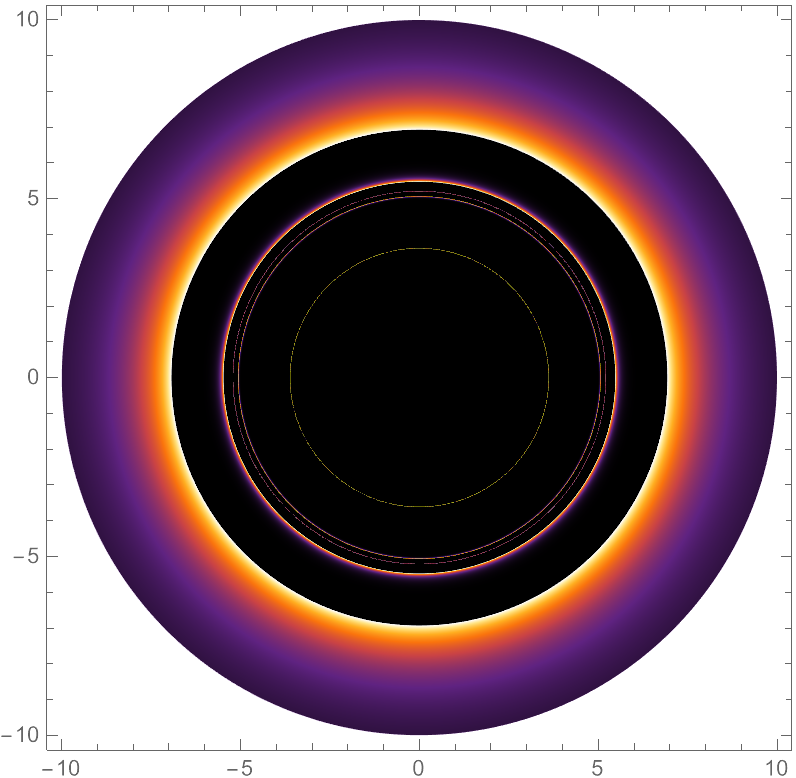}
    \small (b) 
\end{minipage}
\hfill
\begin{minipage}[b]{0.25\textwidth}
    \centering
    \includegraphics[width=\linewidth]{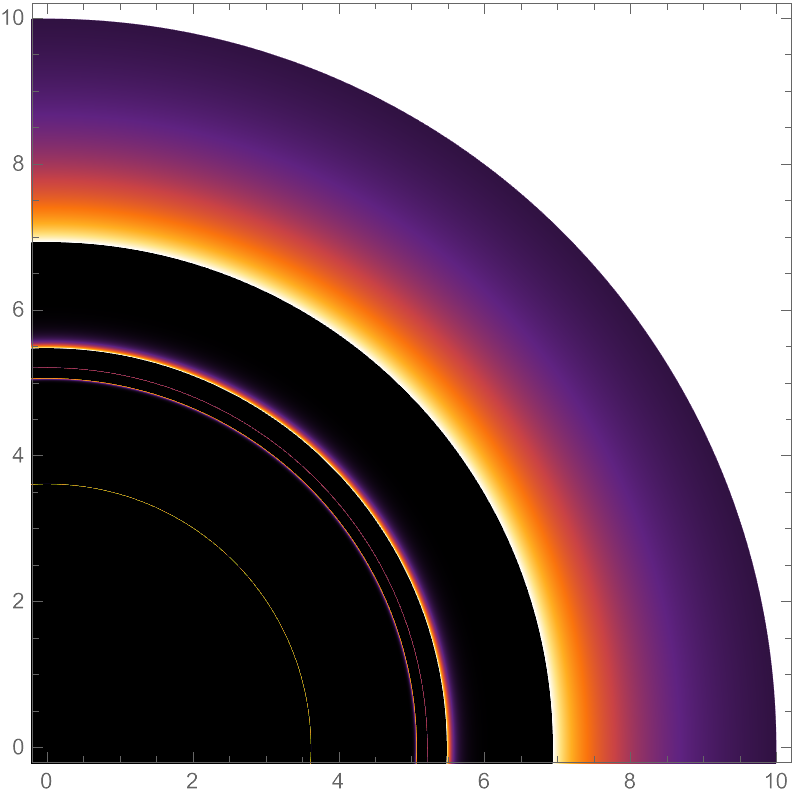}
    \small (c) 
\end{minipage}

\vspace{1mm}  

\begin{minipage}[b]{0.25\textwidth}
    \centering
    \includegraphics[width=\linewidth]{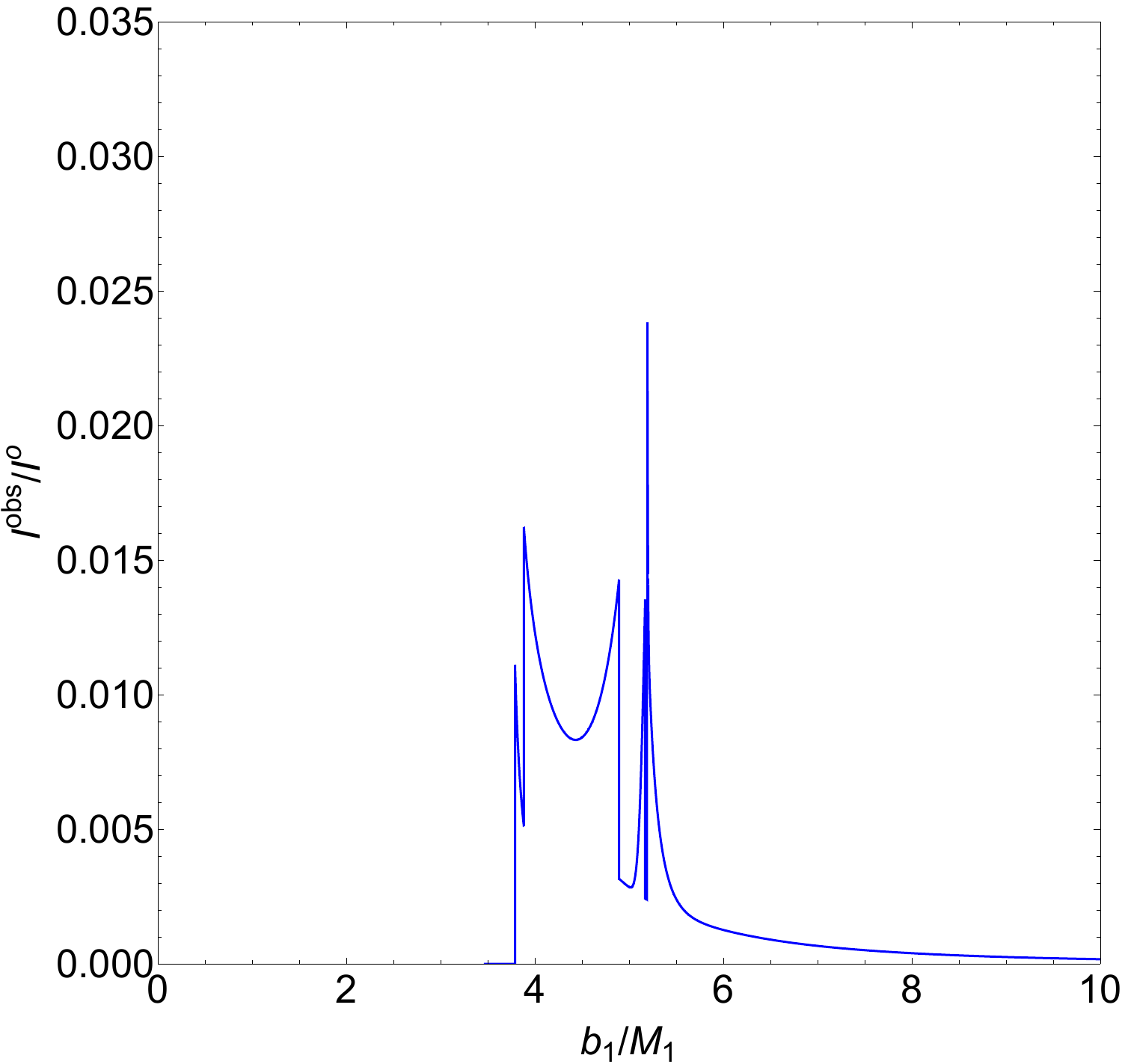}
    \small (d) 
\end{minipage}
\hfill
\begin{minipage}[b]{0.25\textwidth}
    \centering
    \includegraphics[width=\linewidth]{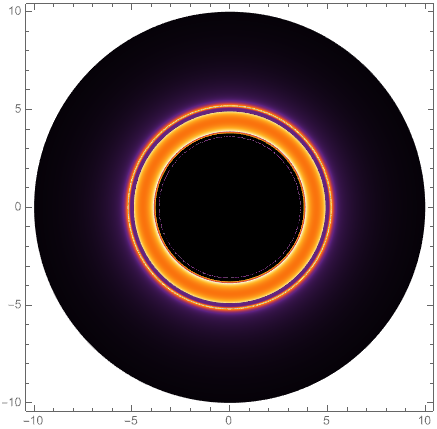}
    \small (e) 
\end{minipage}
\hfill
\begin{minipage}[b]{0.25\textwidth}
    \centering
    \includegraphics[width=\linewidth]{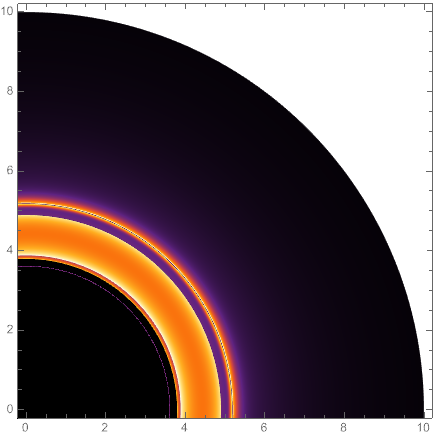}
    \small (f) 
\end{minipage}

\caption{(Color online) Observed intensities (left panels), density plots (middle panels), and local density plots (right panels) for the Schwarzschild ATSW ($\alpha \to 0$) under emission model I (top panels) and emission model II (bottom panels). The parameters are $M_1 = 1$, $M_2 = 1.2$, and $R = 2.6$.}
\label{fig:eight}
\end{figure}

In the special case $\alpha \to 0$ (the Schwarzschild ATSW), the observed intensity profiles, density plots, and local density plots are presented in Fig.~9. The top and bottom rows of Fig.~9 correspond to emission models~I and~II, respectively. For model~I, the direct emission of the Schwarzschild ATSW appears near the critical curve $b_1 \simeq 6.928M_1$, as shown in Fig.~9(a). Two photon rings are located near $b_1 \simeq 5.056M_1$ and $b_1 \simeq 5.210M_1$, whereas the lensing band appears near $b_1 \simeq 5.472M_1$. For model~II, Fig.~9(d) shows that the direct emission is located near $b_1 \simeq 3.4781M_1$. These results indicate that increasing the coupling constant $\alpha$ reduces the size of the light bands outside the shadow, consistent with the 4D EGB black hole case. Nevertheless, Figs.~9(a) and 9(d) show that an extra photon ring (corresponding to the newly introduced second transfer function) exists near $b_1 \simeq 3.600M_1$ for model~I, while an additional lensing band appears near $Z b_{c2} \simeq 3.742M_1$ for model~II. These observations demonstrate that, for the 4D EGB ATSW, the size of these additional light bands increases with $\alpha$, which is opposite to the trend observed for the black hole. Recent work on rotating 4D EGB black holes with thin accretion disks also found that increasing $\alpha$ reduces the shadow size~\cite{Aslam2026}. Hence, the influence of $\alpha$ on these extra rings is opposite to its effect on the corresponding rings in the black hole case.

The reason for this opposite behavior is as follows. In the spacetime $\mathcal{M}_1$, increasing $\alpha$ reduces the horizon radius, the photon-sphere radius, and the critical impact parameter, as summarized in Table~I; the photon ring associated with $\mathcal{M}_1$ therefore shrinks, as in the 4D EGB black hole case. The extra bright rings, however, do not correspond to the photon ring of $\mathcal{M}_1$. They are formed by photons that cross the throat into $\mathcal{M}_2$, are reflected back by the effective potential there, and then return to $\mathcal{M}_1$. Their positions are determined jointly by the critical impact parameters on the two sides, $b_{c_1}$ and $Z b_{c_2}$, rather than by the photon sphere of $\mathcal{M}_1$ alone. As $\alpha$ increases, the gravitational field at a given position in $\mathcal{M}_1$ becomes weaker, so the binding of the returning photons is weakened and the extra rings can form farther out. This explains why the size of the extra rings increases with $\alpha$, in contrast to the black hole case.

\begin{figure}[htbp]
\centering
\begin{minipage}[b]{0.25\textwidth}
    \centering
    \includegraphics[width=\linewidth]{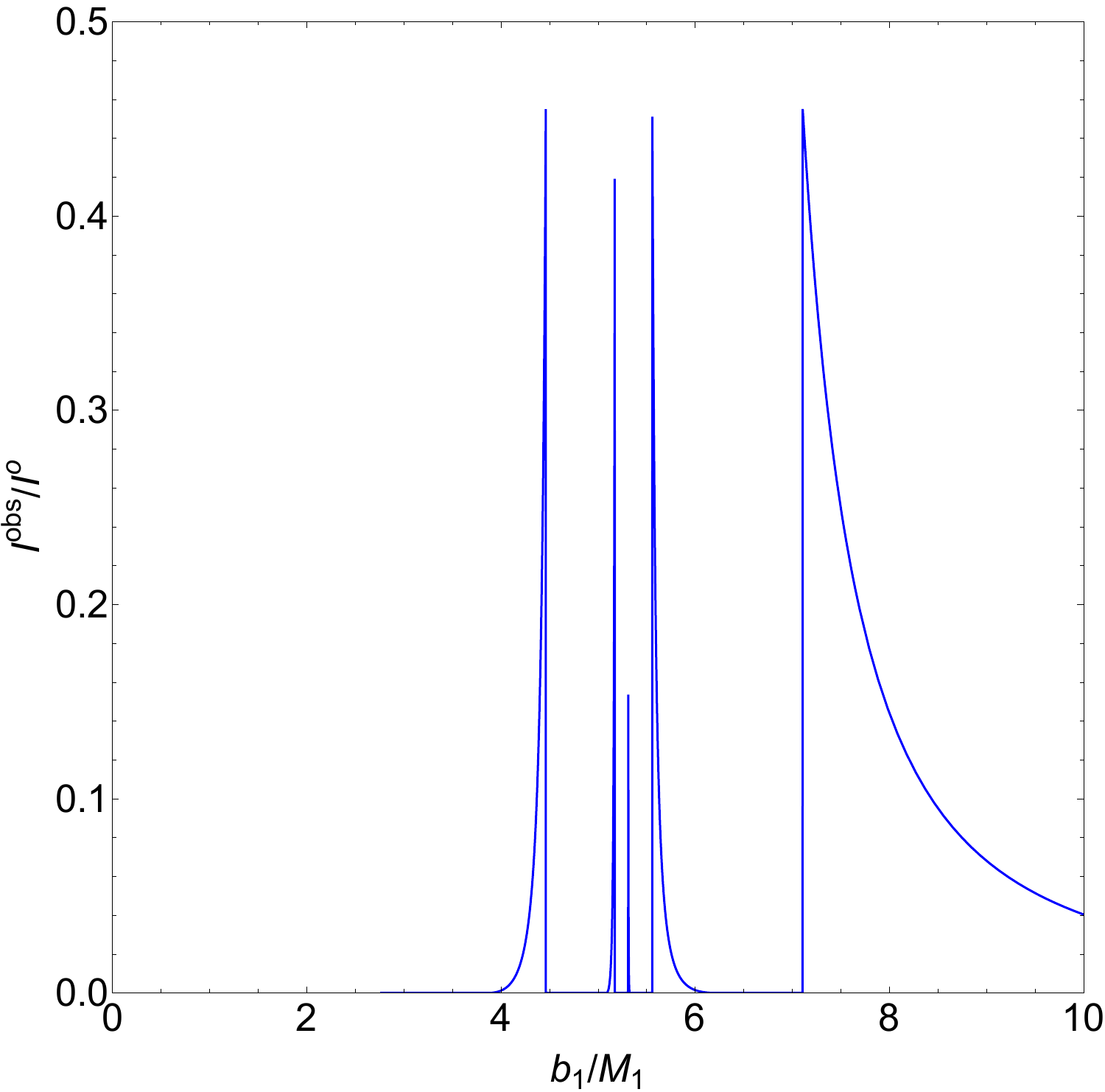}   
    \small (a)  $M_2 = 1.2$, $R = 2.6$  
\end{minipage}
\hfill
\begin{minipage}[b]{0.25\textwidth}
    \centering
    \includegraphics[width=\linewidth]{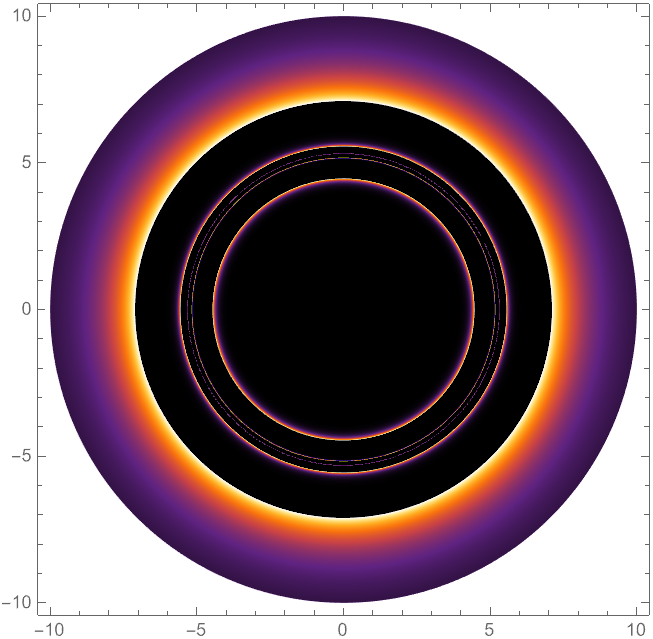}
    \small (b) $M_2 = 1.2$, $R = 2.6$
\end{minipage}
\hfill
\begin{minipage}[b]{0.25\textwidth}
    \centering
    \includegraphics[width=\linewidth]{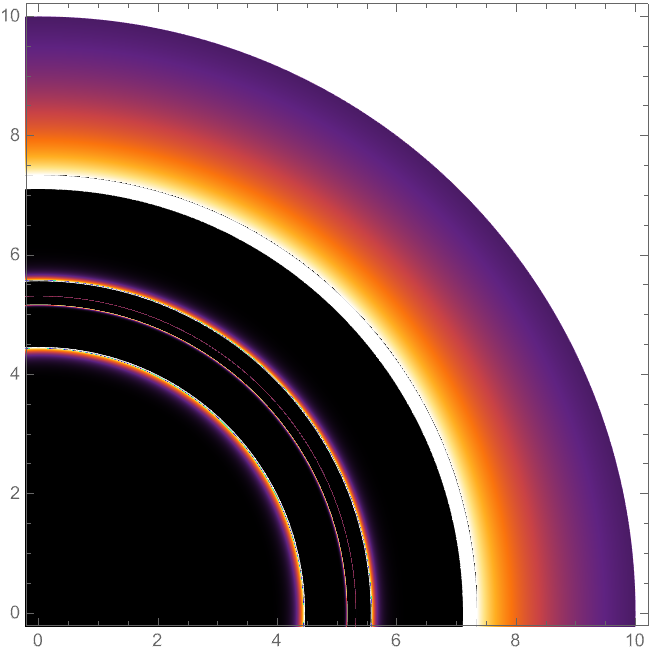}
    \small (c) $M_2 = 1.2$, $R = 2.6$
\end{minipage}

\vspace{1mm}  

\begin{minipage}[b]{0.25\textwidth}
    \centering
    \includegraphics[width=\linewidth]{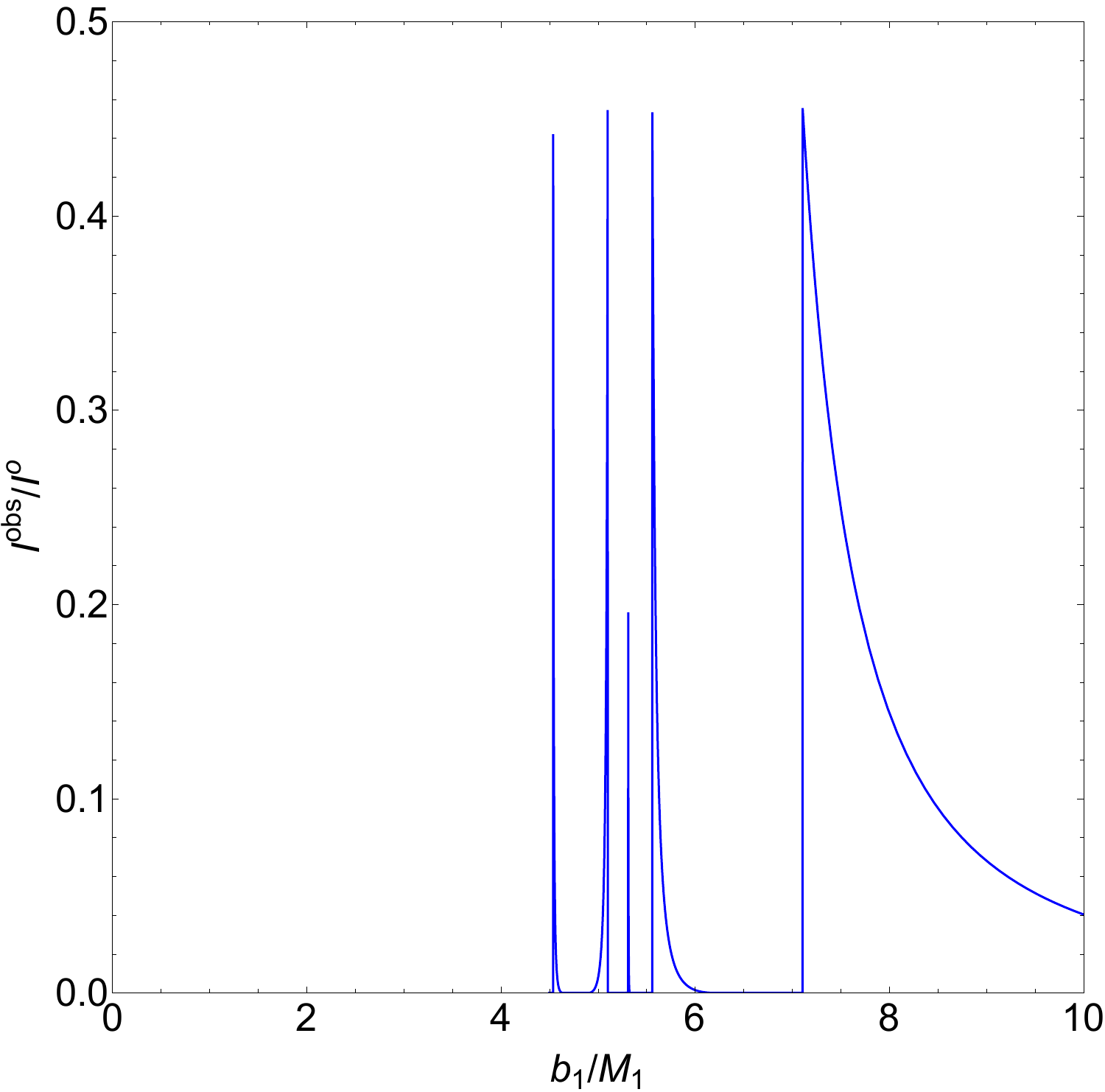}
    \small (d)  $M_2 = 1.1$, $R = 2.6$
\end{minipage}
\hfill
\begin{minipage}[b]{0.25\textwidth}
    \centering
    \includegraphics[width=\linewidth]{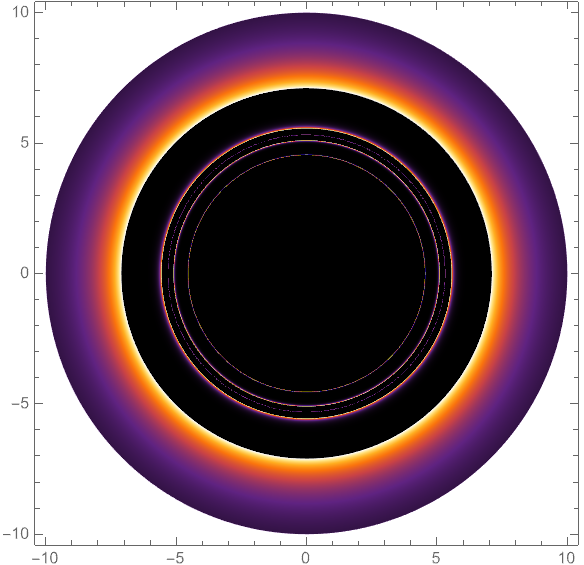}
    \small (e) $M_2 = 1.1$, $R = 2.6$
\end{minipage}
\hfill
\begin{minipage}[b]{0.25\textwidth}
    \centering
    \includegraphics[width=\linewidth]{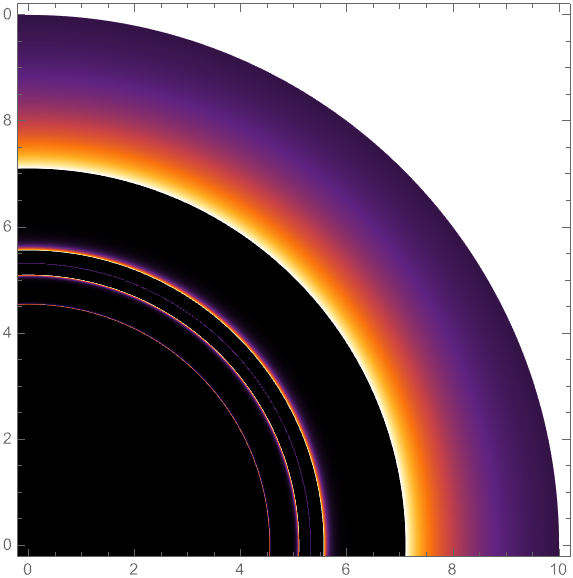}
    \small (f) $M_2 = 1.1$, $R = 2.6$
\end{minipage}

\vspace{1mm}

\begin{minipage}[b]{0.25\textwidth}
    \centering
    \includegraphics[width=\linewidth]{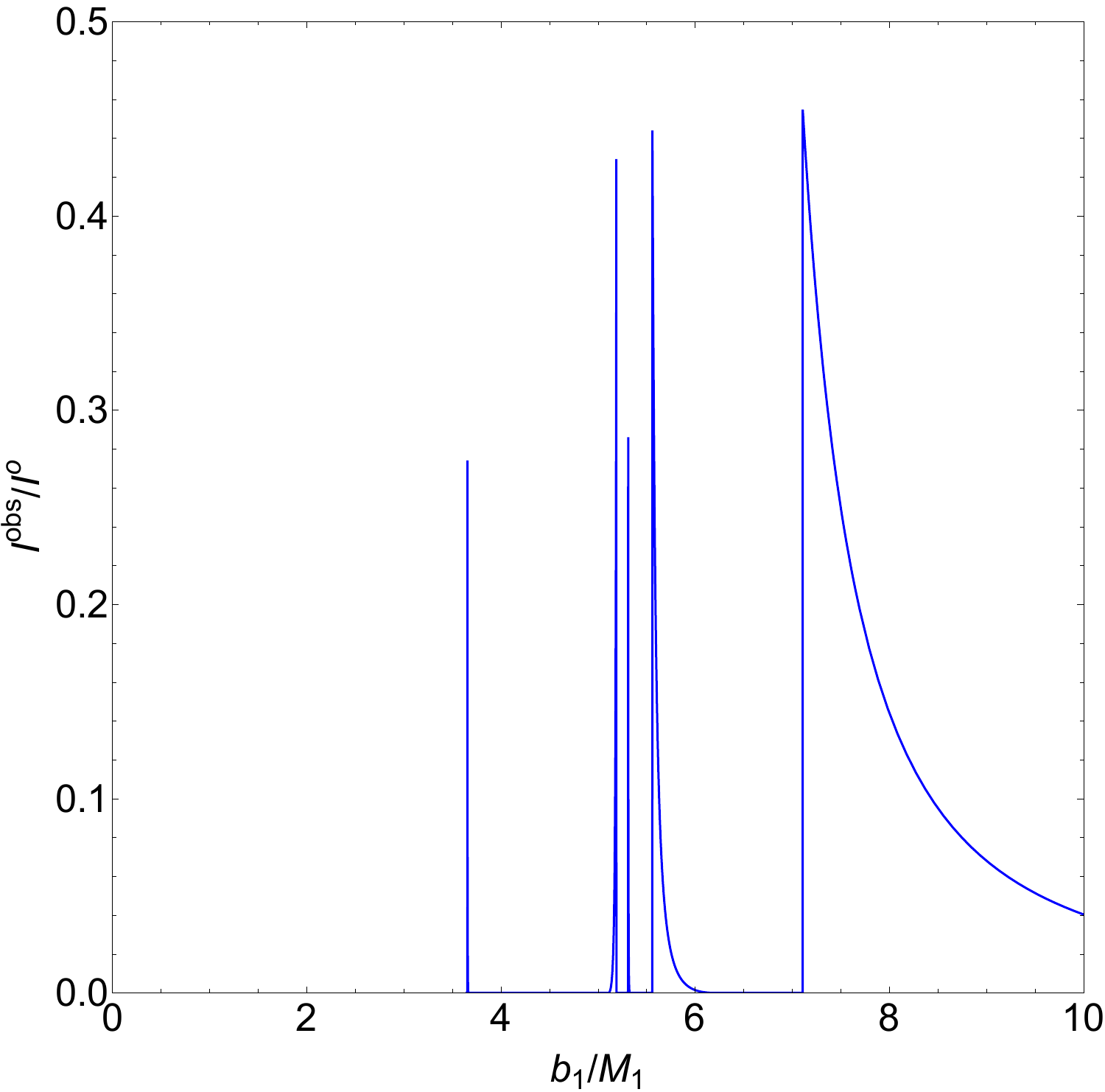}
    \small (g) $M_2 = 1.2$, $R = 2.7$
\end{minipage}
\hfill
\begin{minipage}[b]{0.25\textwidth}
    \centering
    \includegraphics[width=\linewidth]{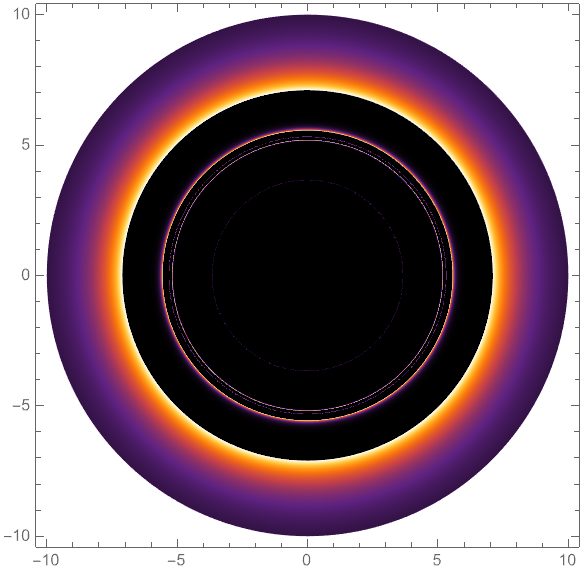}
    \small (h) $M_2 = 1.2$, $R = 2.7$
\end{minipage}
\hfill
\begin{minipage}[b]{0.25\textwidth}
    \centering
    \includegraphics[width=\linewidth]{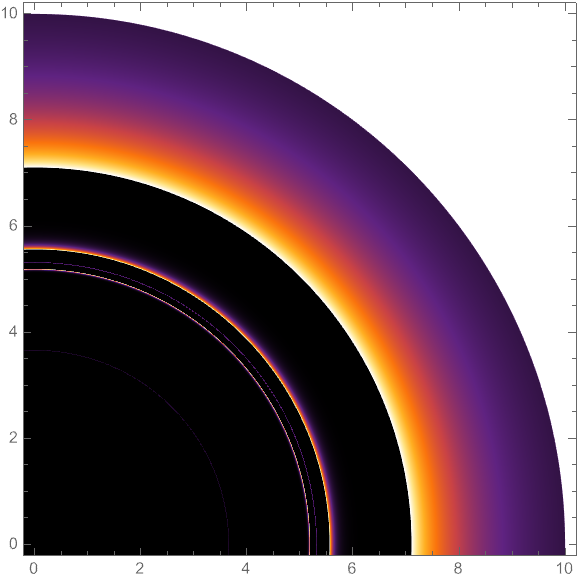}
    \small (i) $M_2 = 1.2$, $R = 2.7$
\end{minipage}

\caption{(Color online) Emission Model I: observed intensities (left panel), density plots (middle panel), and local density plots (right panel) of the ATSW for $\alpha = -0.3$. }  
\label{fig:ten}
\end{figure}

We further investigate how variations in the mass ratio $M_2/M_1$ and the throat radius $R$ affect the optical characteristics of ATSWs. To isolate the influence of the mass ratio, we fix $M_1 = 1$ and vary $M_2$, with $\alpha = -0.3$ in all cases to ensure consistent comparisons across different mass ratios. A comparison of Fig.~10(a) with Fig.~10(d) shows that, as the mass ratio decreases, the radius of the innermost extra photon ring increases, whereas the radius of the second extra photon ring remains nearly unchanged. Consequently, the separation between the two extra photon rings gradually decreases. This trend is particularly evident in Figs.~10(c) and~10(f).

We then fix $\alpha = -0.3$ and $M_2 = 1.2$ and vary the throat radius $R$. A comparison of Fig.~10(a) with Fig.~10(g) reveals that, as $R$ increases, the radius of the innermost extra photon ring decreases, whereas the radius of the second extra photon ring remains nearly unchanged. Consequently, the separation between the two extra photon rings gradually increases. This trend is particularly evident in Figs.~10(c) and~10(i). Our findings on the effects of the mass ratio and throat radius are consistent with the results for non-commutative ATSWs reported in Ref.~\cite{Wu2025}.

\section{Conclusion and Discussion}\label{sec4}

To study the optical appearance of the 4D EGB asymmetric thin-shell wormhole, we employed Visser's cut-and-paste construction to join two spacetimes, $\mathcal{M}_1$ and $\mathcal{M}_2$. A static observer is assumed to reside in $\mathcal{M}_1$. We computed the photon sphere radius $r_{ph}$ and the critical impact parameter $b_c$ for various values of the Gauss-Bonnet coupling constant $\alpha$. The results show that increasing $\alpha$ reduces both $r_{ph}$ and $b_c$; that is, a larger $\alpha$ shifts the photon sphere inward.

The effective potentials of the wormhole and the 4D EGB black hole are presented in Fig.~2. Depending on the impact parameter, the photon trajectory belongs to one of three classes: (i) $b_1 < Z b_{c_2}$, for which the photon crosses into $\mathcal{M}_2$ and escapes to infinity there; (ii) $Z b_{c_2} < b_1 < b_{c_1}$, for which the photon enters $\mathcal{M}_2$, turns back, and returns to $\mathcal{M}_1$; and (iii) $b_1 > b_{c_1}$, for which the photon remains in $\mathcal{M}_1$ and is reflected back to infinity. Fig.~3 displays the corresponding trajectories. As $b_1$ decreases, the path inside $\mathcal{M}_2$ becomes longer. Moreover, increasing $\alpha$ decreases $b_{c_1}$ and increases $Z b_{c_2}$. These trends are consistent with Tab.~I and Fig.~3.

We studied the orbit numbers and transfer functions, identifying new second and third transfer functions corresponding to a ``lensing band'' and a ``photon ring group'', respectively. Using two emission models for the thin accretion disk, we compared the observational appearances of the wormhole and the black hole. In emission model~I, two additional photon rings appear near the critical curves $b_1 \simeq 4.457M_1$ and $b_1 \simeq 5.169M_1$. In emission model~II, an additional lensing band is observed between $Z b_{c_2} \simeq 2.766M_1$ and $b_{c_1} \simeq 5.30178M_1$. A comparison between the Schwarzschild wormhole (the limit $\alpha \to 0$) and the 4D EGB wormhole shows that the size of these additional light bands increases with $\alpha$. This behavior is opposite to that observed for a black hole, where increasing $\alpha$ shrinks the rings.

This opposite trend arises because the extra rings are controlled jointly by $b_{c_1}$ and $Z b_{c_2}$, the latter of which also depends on $\alpha$. Consequently, the response of the extra rings to $\alpha$ differs from that of ordinary black-hole rings. Hence, the distinct trend of the extra rings may provide a potentially useful observational criterion for distinguishing the 4D EGB asymmetric thin-shell wormhole from a black hole, although a more realistic treatment of the accretion flow is needed to confirm this.

In addition to the 4D EGB ATSW shown in the upper row, the lower row of Fig.~\ref{fig:three} presents the transfer functions of three comparison models with the same parameter values: (d) the Schwarzschild ATSW, (e) the Hayward ATSW with $g=0.3$, and (f) the Kalb-Ramond ATSW with $l=0$ and $Q=0.3$, where $M_1=1$, $M_2=1.2$, and $R=2.6$. A comparison of panels (c), (e), and (f) shows clear differences in the extra transfer functions. The extra transfer functions appear farthest to the right for the 4D EGB ATSW and farthest to the left for the Kalb-Ramond ATSW, with the Hayward ATSW in between. The slope of the third transfer function is steepest for the 4D EGB ATSW and least steep for the Kalb-Ramond ATSW, resulting in different luminosities of the third photon ring. The range of $b_1$ covered by the extra transfer functions is widest for the Kalb-Ramond ATSW and narrowest for the 4D EGB ATSW, and the separation between the dashed and solid curves follows the same order.

\begin{figure}[htbp]
\centering
\begin{minipage}[b]{0.25\textwidth}
    \centering
    \includegraphics[width=\linewidth]{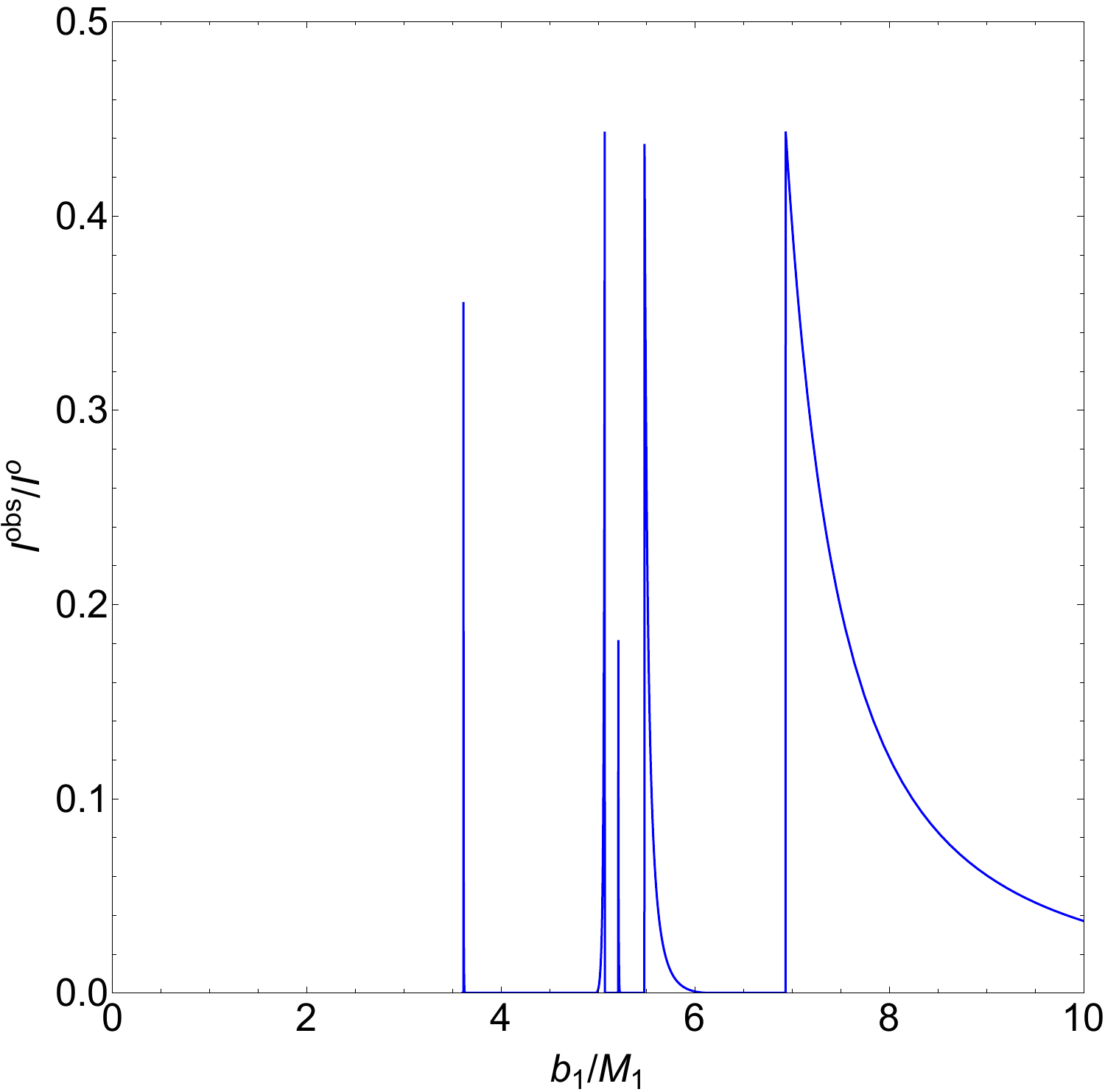}   
    \small {(a)}    
\end{minipage}
\hfill
\begin{minipage}[b]{0.25\textwidth}
    \centering
    \includegraphics[width=\linewidth]{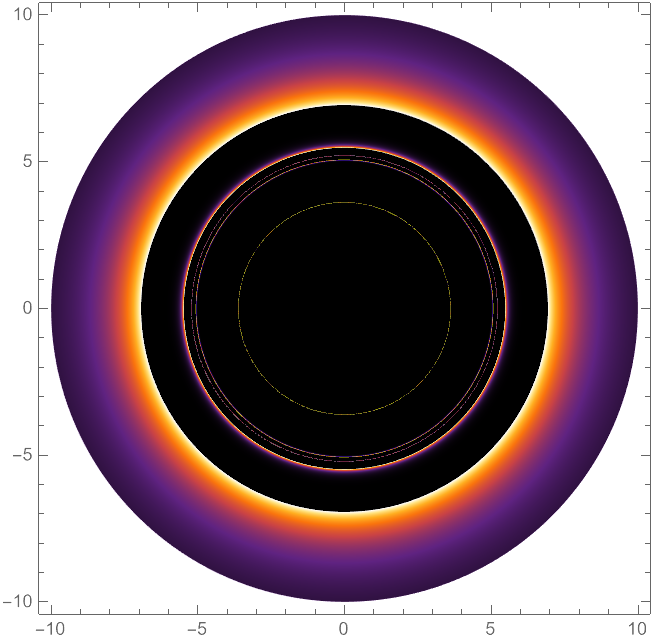}
    \small {(b)}
\end{minipage}
\hfill
\begin{minipage}[b]{0.25\textwidth}
    \centering
    \includegraphics[width=\linewidth]{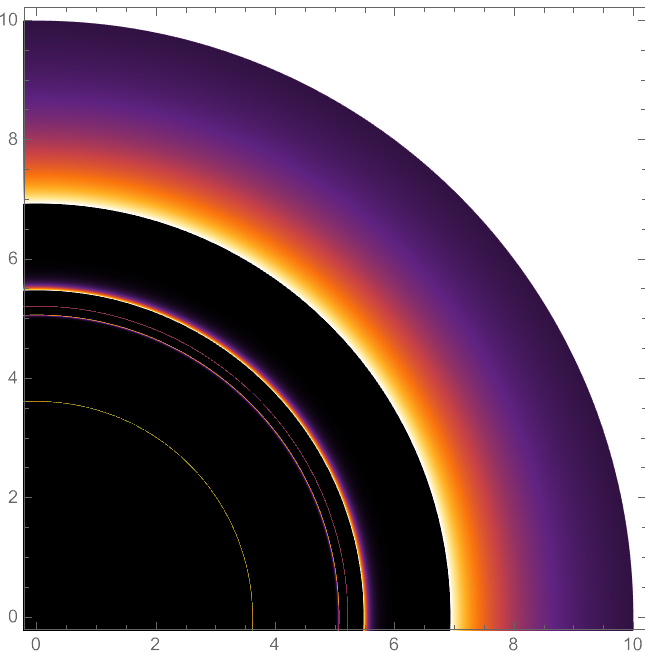}
    \small {(c)}
\end{minipage}

\vspace{1mm}  

\begin{minipage}[b]{0.25\textwidth}
    \centering
    \includegraphics[width=\linewidth]{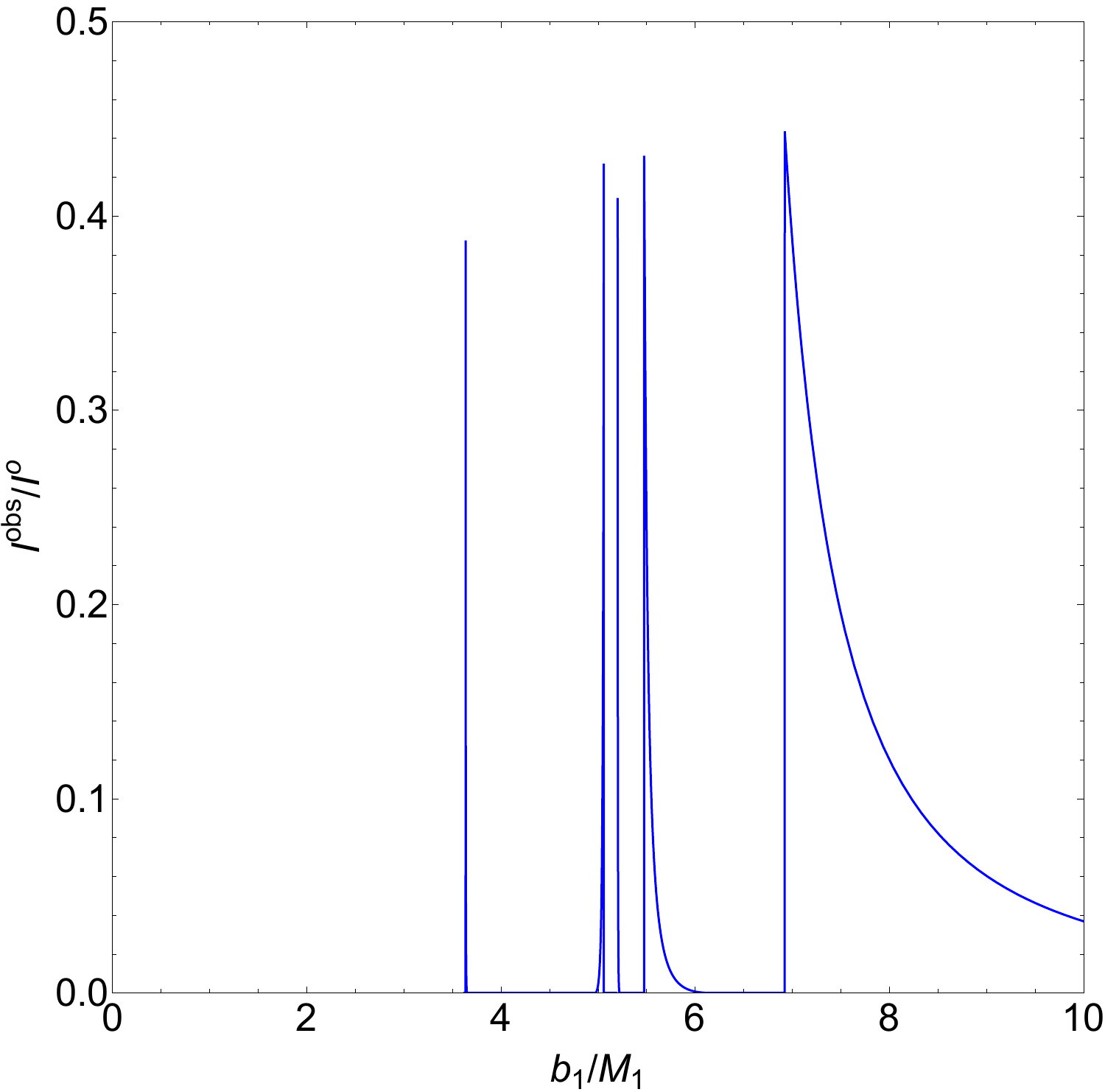}
    \small  {(d)}
\end{minipage}
\hfill
\begin{minipage}[b]{0.25\textwidth}
    \centering
    \includegraphics[width=\linewidth]{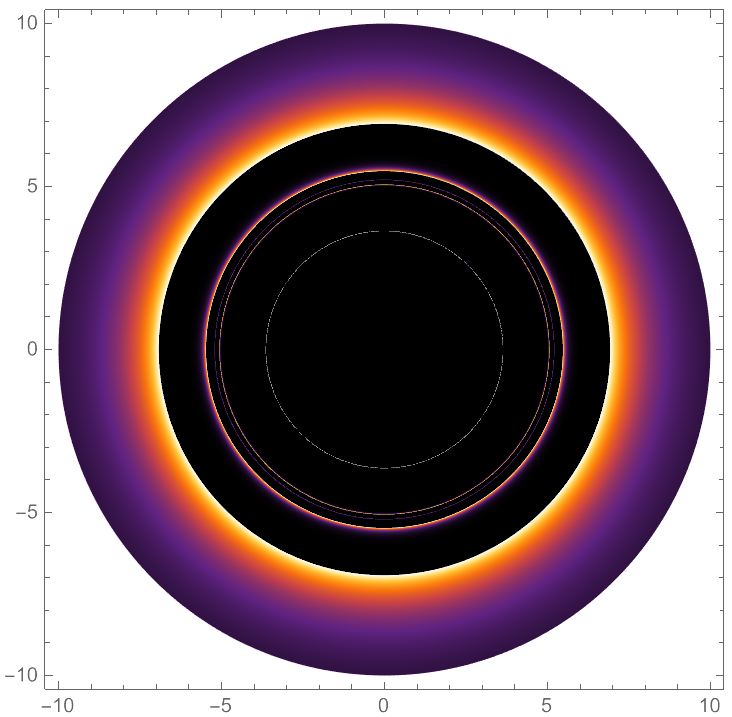}
    \small   {(e)}
\end{minipage}
\hfill
\begin{minipage}[b]{0.25\textwidth}
    \centering
    \includegraphics[width=\linewidth]{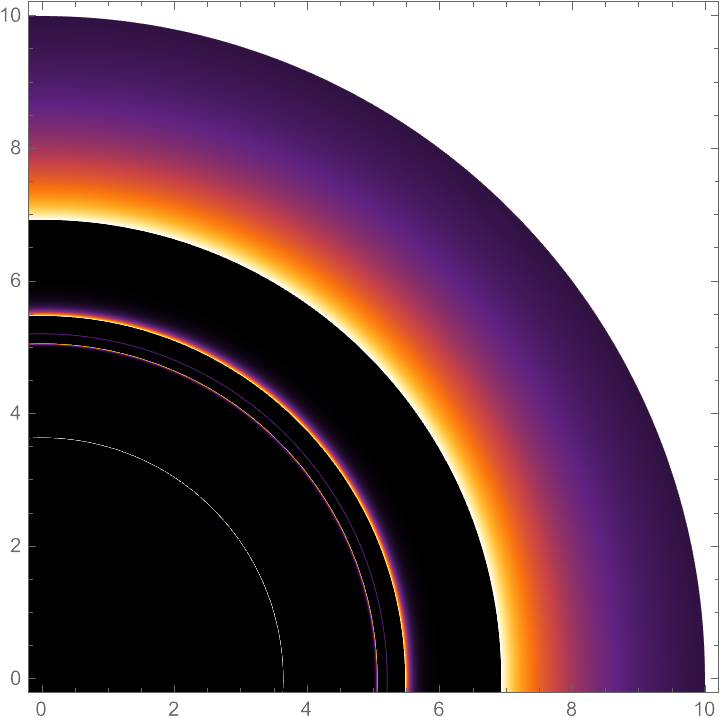}
    \small  {(f)}
\end{minipage}

\vspace{1mm}  

\begin{minipage}[b]{0.25\textwidth}
    \centering
    \includegraphics[width=\linewidth]{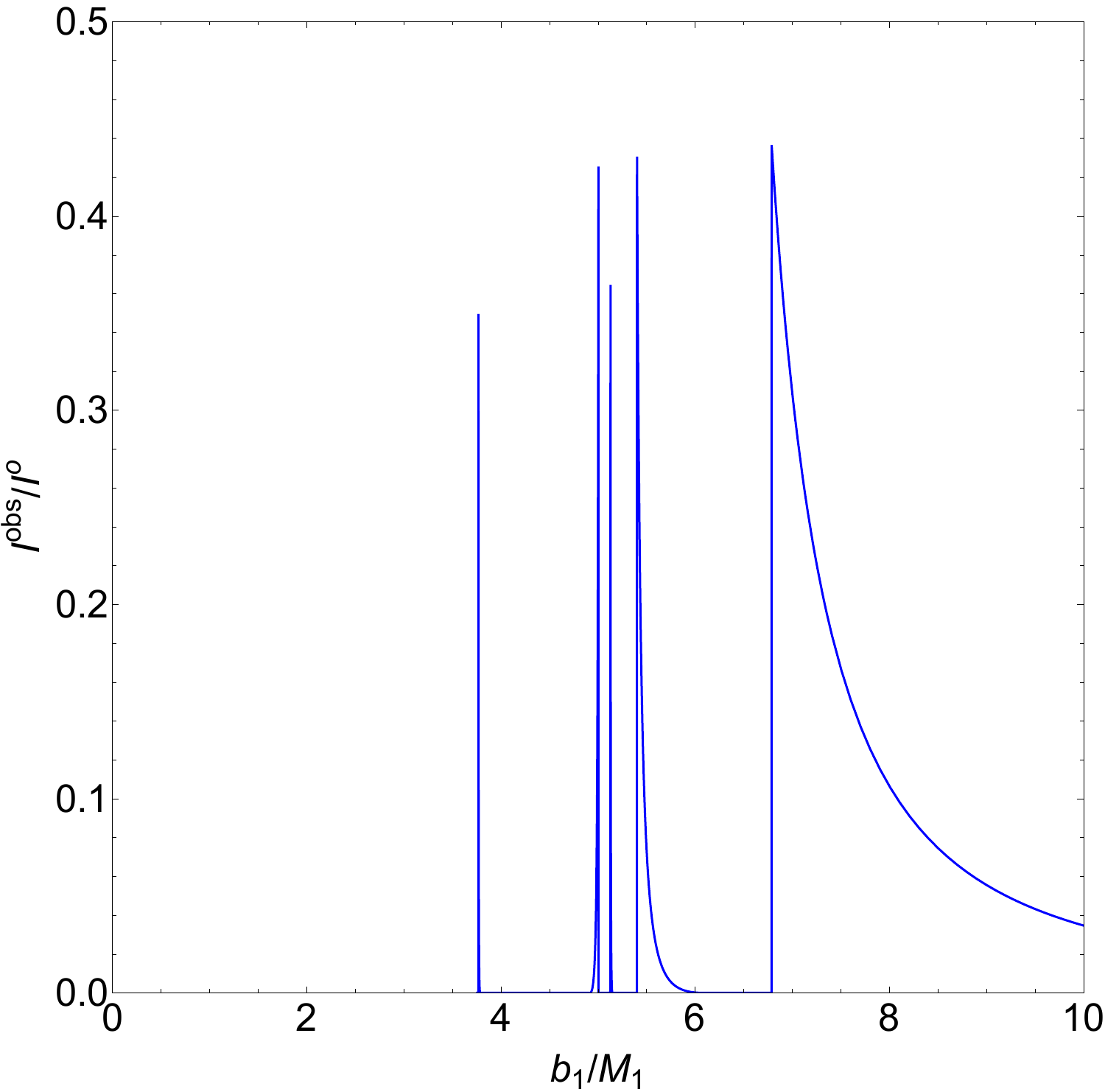}
    \small {(g)}
\end{minipage}
\hfill
\begin{minipage}[b]{0.25\textwidth}
    \centering
    \includegraphics[width=\linewidth]{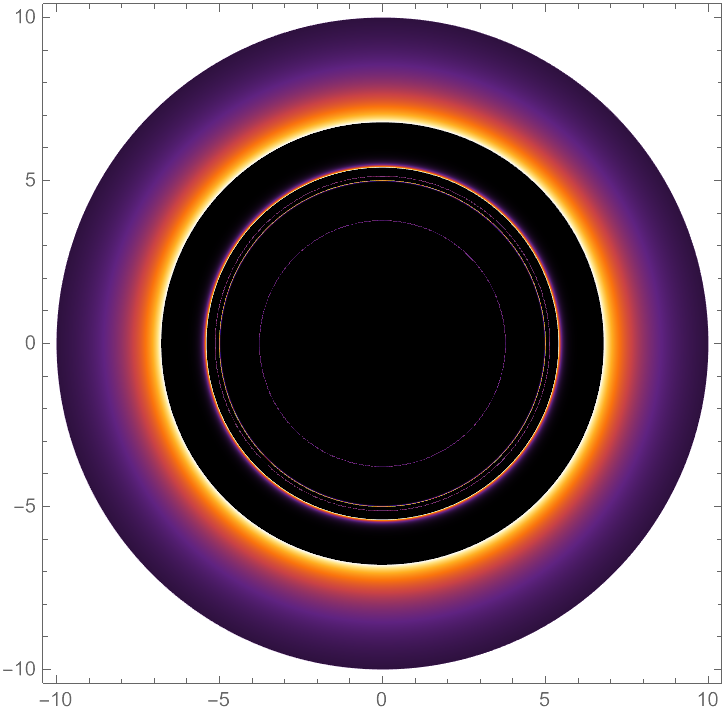}
    \small  {(h)}
\end{minipage}
\hfill
\begin{minipage}[b]{0.25\textwidth}
    \centering
    \includegraphics[width=\linewidth]{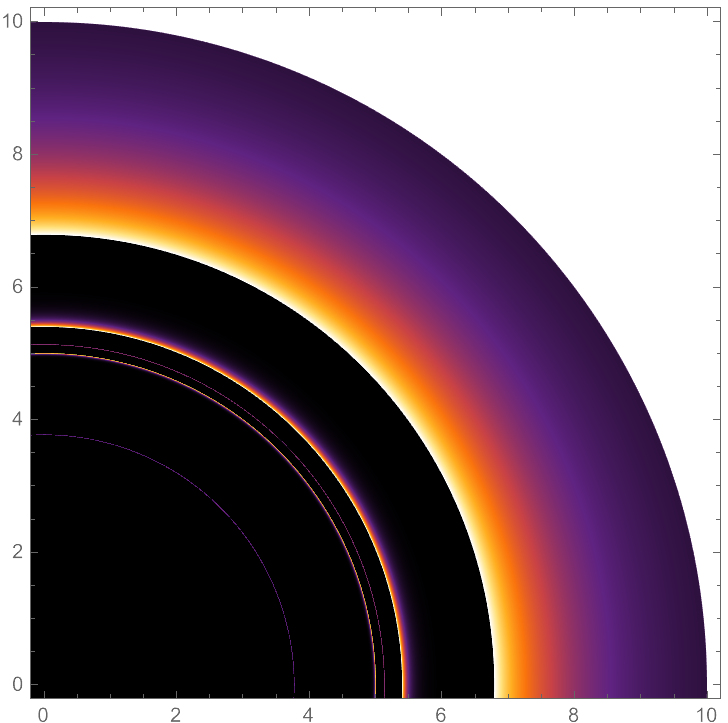}
    \small {(i)}
\end{minipage}

\caption{(Color online) Observed intensities (left panels), density plots (middle panels), and local density plots (right panels) under emission model I for three comparison models: the Schwarzschild ATSW (top row), the Hayward ATSW with $g=0.3$ (middle row), and the Kalb-Ramond ATSW with $l=0$ and $Q=0.3$ (bottom row). The remaining parameters are fixed at $M_1=1$, $M_2=1.2$, and $R=2.6$.}
\label{fig:11}
\end{figure}

We also examined how the mass ratio $M_2/M_1$ and the throat radius $R$ affect the optical appearance of ATSWs. For a fixed $\alpha = -0.3$, a lower mass ratio enlarges the innermost extra photon ring while leaving the second extra ring unchanged, thereby reducing the ring separation. Conversely, increasing the throat radius shrinks the innermost extra ring, whereas the second extra ring remains nearly constant, leading to a larger gap. These trends are clearly shown in the figures. Accordingly, these observational signatures could provide a potential criterion for distinguishing ATSWs from black holes in future astronomical observations, provided that more realistic accretion models are considered.
\section*{Acknowledgments}
  This work was supported in part by the National Natural Science Foundation of China under Grant No. 12675078 and by the Basic Scientific Research Operating Expenses Program of China West Normal University under Grant No. 2026kx007.

\end{document}